\documentstyle[12pt,epsf]{article}
\newcommand{\bbox}{\vec}
\input{commands}
\begin{document}
\begin{titlepage}

\newcommand\letterhead {%
\hfill\parbox{8cm}{    \large \it UNIVERSITY of PENNSYLVANIA\\
\large Department of Physics\\
David Rittenhouse Laboratory\\
Philadelphia PA 19104--6396}\\[0.5cm]
{\bf PREPRINT UPR--0128MT}\\ July 1993}
\noindent\letterhead\par
\vspace{2 cm}
\noindent
\begin{center}
{\Large \bf Inertial parameters of the Skyrmion-Skyrmion system with the
product ansatz}\\
\end{center}
\begin{center}
{\bf \large
Bin Shao, Niels R. Walet,~and R. D. Amado}
\end{center}
\begin{center}
\vfill
{\em Submitted to Physical Review C}
\end{center}
\end{titlepage}

\title{Inertial parameters of the Skyrmion-Skyrmion system with the
product ansatz}

\author{Bin Shao, Niels R. Walet\thanks{address after Sept.~1, 1993:
Instit\"ut f\"ur theoretische Physik III, Universit\"at Erlangen-N\"urnberg,
D-91058 Erlangen, Germany},~and R. D. Amado\\
Department of Physics,\\
 University of Pennsylvania,\\
Philadelphia, PA 19104}

\maketitle

\begin{abstract}
The kinetic energy of the Skyrmion-Skyrmion system is
calculated in the product ansatz. The energy is expanded in a
set of rotationally invariant quadratic functions
of translational and iso-rotational velocities in the product
ansatz and all  inertial parameters are evaluated.
Most of the iso-spin
dependent parameters are small  at separations larger than 1 fm.
However, some parameters of the coupling between the iso-rotational
velocities of the
Skyrmions decay slowly, as $1/R$, when zero pion mass
is used. Otherwise we find that the results are not sensitive to the
addition of the pion mass term.\\[0.3cm]
\end{abstract}


\section{Introduction}
\label{sec:intro}

The Skyrme model is a classical non-linear field theory of interest
both for itself and because it is a model of QCD in the limit of a
large number of colors ($N_c$) \cite{Skyrme}. It has been studied extensively
in the baryon number one and number two sectors \cite{B1sector,B2sector,%
Bible,WA}.
Recent results for baryon number two ($B=2$) \cite{WA}
have shown that careful treatment
of the non-linear nature of the equations and of quantum corrections
(finite $N_c$ effects), lead to a very good description of the
nucleon-nucleon static potential.  One finds one pion exchange,
mid-range attraction, a tensor force and short range repulsion, all
in substantial agreement with phenomenological potentials, and
even with phase shifts \cite{Wphaseshift}.
The next step is the consideration of non-static
effects.  Here there has been less work \cite{Oka,Yazaki,Riska,Otofuji}
and no systematic study.

In this paper we present the analysis and results of a systematic study of
the kinetic energy in the $B=2$ Skyrmion system.
We do so in the context of the product ansatz. This has the advantage of
relative algebraic simplicity and of explicit identifiability of the
collective coordinates.  Its main disadvantage is that it is a poor
approximation to the $B=2$ exact solution at small Skyrmion separation.
For example, the product ansatz cannot model the lowest energy $B=2$
static solution of the Skyrme Lagrangian in which the baryon density has
toroidal symmetry and the identity of the individual Skyrmions is
completely lost.  For the static nucleon-nucleon interaction, the
product ansatz is particularly dangerous because that interaction is
extracted by subtracting two large numbers to get a relatively small
one, and even moderate errors in configurations get greatly
magnified.  That is why finding the mid-range attraction between nucleons
waited for a full solution of the non-linear equations.
For the non-static terms there is no corresponding subtraction and we might
expect the product ansatz to be more reliable.  However even here it
is far from perfect.  For example we find kinetic energy terms that
are embarrassingly large, yet violate parity and
particle interchange symmetry.
These are artifacts of this ansatz, that has both these diseases.  Such
terms would not be present in an exact calculation.  Nevertheless, as a
first orientation, we calculate in the product ansatz.

In the product ansatz, the $B=2$ Skyrmion system is represented by
two $B=1$ Skyrmions separated by a distance $\bbox{R}$.  Each of these
two Skyrmions can also be rotated in iso-spin space.  For the static
system, the energy of the $B=2$ state depends on $\bbox{R}$ and on the
relative iso-spin orientation of the two Skyrmions.  To study non-static
effects, we make each of these three collective
vector coordinates, $\bbox{R}$,
and the iso-angle of each Skyrmion, time dependent.
Now the kinetic energy can depend on any combination of the
relative Skyrmion velocity and the two angular velocities.  Clearly
there are no terms in the kinetic energy linear in velocity.
Since the Skyrme Lagrangian is quadratic in time derivatives, the
kinetic energy is only bilinear in the velocities.  In principle
all possible terms connecting velocities can occur in the kinetic
energy, and since there are three vector velocities, the mass matrix
will be nine by nine. The elements of this mass matrix depend on
$\bbox{R}$ (both its magnitude and direction), and the relative iso-spin
orientation of the two Skyrmions, and are calculated as integrals over
the single Skyrmion profile functions.
There are many terms in the expansion of the kinetic energy in
the relative iso-orientation.  We find that terms have a steep hierarchy
in size with relative iso-orientation, the terms that are independent of
orientation being the largest. We also find, as mentioned above, that not
all terms that violate parity and exchange symmetry are zero, due to the
faults of the product ansatz.  There are many terms in our expansion.
In this paper we present the algebraic form for all of them but
only discuss the largest ones.

After the mass matrix has been constructed, it has to be inverted
to pass from a velocity to a canonical momentum form so that a
quantum mechanical Hamiltonian can be constructed.  We do
not carry out that step here and only mention that
attention to the requirements of matrix inversion are called for.
We have recently demonstrated \cite{SpinOrbit}
the role that the transformation from
velocities to momenta plays in getting the correct sign
for the nucleon-nucleon spin-orbit interaction.

In Sec.~\ref{sec:2} we define notation, and substitute the $B=2$ product
ansatz with time dependent collective coordinates in the Skyrme Lagrangian.
We then develop the expansion of the terms in the kinetic energy.  In
Sec.~\ref{sec:3} we present the algebraic structure of the results of that
expansion.  Section \ref{sec:4} discusses and presents graphically the
 behavior
of some of the leading terms in our expansion.
Section \ref{sec:5} presents some conclusions. Finally
some technical details are given in the Appendix.  The main purpose
of this paper is to present the full expansion of the kinetic energy
both in systematic algebraic form, and as calculated results.  We
leave the use of these forms and results in particular physics problems
to further publications.

\section{Skyrme Model}
\label{sec:2}

The Skyrme model Lagrange density can be written as \cite{Skyrme}
\begin{eqnarray}
{\cal L}&=&-\frac{f_{\pi}^{2}}{4}\mbox{Tr}L_{\mu}L_{\mu}
+\frac{1}{32e^{2}}\mbox{Tr}[L_{\mu},L_{\nu}]^{2}
\nonumber\\&&+\frac{f_{\pi}^2 m_{\pi}^2}{4}
[\mbox{Tr}(U+U^{\dagger})-4],  \label{eq:Lag}
\end{eqnarray}
where $L_{\mu}=\partial_{\mu}U U^{\dagger}$ is the left Sugawara
variable for the meson field $U$ which is a unitary $SU(2)$ matrix.
There are only three parameters in the Lagrangian, $m_{\pi}$, the pion
mass, $f_{\pi}$, the pion decay constant, and $e^2$, a dimensionless
constant normally adjusted to give the correct nucleon-delta mass splitting.
It is worth noting that the Skyrme Lagrangian is quadratic in
time derivatives. The model has a topologically conserved quantum number that
Skyrme identified as the baryon number, $B$.
Since the equations of motion derived from
(\ref{eq:Lag}) are highly nonlinear, it is not possible, in general, to
obtain exact analytical solutions. The standard time-independent solution to
the classical field equations
for $B=1$ is the defensive hedgehog, where the pion field points radially
outward,
\begin{equation}
U(\bbox{r}) = \exp( i\bbox{\tau} \cdot \hat{r} F(r) ).   \label{eq:profile}
\end{equation}
The baryon number of this state is given by $B=(F(0)-F(\infty))/\pi=1$.
This solution breaks translational invariance, as well as the $O(4)$
spin-iso-spin symmetry. If we perform a global SU(2) iso-rotation on the state,
\begin{equation}
U(\bbox{r}|A) = A^\dagger U(\bbox{r}) A,
\end{equation}
we obtain a state of the same energy.

For the $B=2$ system,
the product ansatz in which $U$ is written as a product of
two single Skyrmions is often used. The product ansatz is a poor
approximation to the exact solution
at short distances but it is expected to be valid
when the two Skyrmions are far apart.
The advantage of using this ansatz is that, with it,
the collective coordinates
can be easily identified.
We write $U$ for the Skyrmion-Skyrmion system as
\begin{equation}
U=\tilde{U}_{1}\tilde{U}_{2},  \label{eq:Prdtansatz}
\end{equation}
with
\begin{equation}
\tilde{U}_{1}=A_{1}(t)U_{1}(\bbox{x}_{1})A_{1}^{\dagger}(t),
\end{equation}
and
\begin{equation}
\tilde{U}_{2}=A_{2}(t)U_{2}(\bbox{x}_{2})A_{2}^{\dagger}(t),
\end{equation}
where $U_{1}$ and $U_{2}$ are both single Skyrmion solutions,
$\bbox{x}_{1}=\bbox{r}-{\bbox{R}}/{2}$ and $\bbox{x}_{2}=\bbox{r}
+{\bbox{R}}/{2}$
are the positions of the two Skyrmions with a relative distance R.
Finally $A_{1}(t)$ and $A_{2}(t)$ are the $SU(2)$ iso-rotation matrices
for the two Skyrmions.

It is often convenient to work with the so-called Sugawara variables of
a $SU(2)$ matrix $A$ (which can be $U$, $\tilde{U}_{1}$,$\tilde{U}_{2}$,
$U_{1}$, $U_{2}$ , $A_{1}$ or $A_{2}$)
instead of the matrix itself.
The left Sugawara variable is defined as
\begin{equation}
L_{\mu}(A)=(\partial_{\mu}A) A^{\dagger},
\end{equation}
while the right one is
\begin{equation}
R_{\mu}(A)=A^{\dagger}\partial_{\mu}A.
\end{equation}
The two sets of the Sugawara variables are related,
\begin{equation}
L_{\mu}(A)=AR_{\mu}(A)A^{\dagger}.
\end{equation}
It can be shown that both $L_{\mu}$ and $R_{\mu}$ are traceless.
We can, therefore,
express these variables in terms of their three independent components
\begin{equation}
L_{\mu}(A)=i\tau^{a}L^{a}_{\mu}(A),
\end{equation}
and
\begin{equation}
R_{\mu}(A)=i\tau^{a}R^{a}_{\mu}(A).
\end{equation}

It is easy to show that for $U$ given in (\ref{eq:Prdtansatz}):
\begin{equation}
L_{\mu}(U)=\tilde{U}_{1}[R_{\mu}(\tilde{U}_{1})+L_{\mu}(\tilde{U}_{2})]
\tilde{U}_{1}^{\dagger}.
\end{equation}
Using the trace formulas $\mbox{Tr}(\bbox{\tau}\cdot\bbox{a}\,
\bbox{\tau}\cdot\bbox{b})
=2\bbox{a}\cdot\bbox{b}$ and $\mbox{Tr}\left[ [\bbox{\tau}\cdot\bbox{a},
\bbox{\tau}\cdot\bbox{b}][\bbox{\tau}\cdot\bbox{c},\bbox{\tau}\cdot\bbox{d}]
\right]
=8(\bbox{a}\cdot\bbox{d})(\bbox{b}\cdot\bbox{c})-
8(\bbox{a}\cdot\bbox{c})(\bbox{b}\cdot\bbox{d})$, we find
 for the kinetic energy density
\begin{equation}
{\cal K}=\frac{f_{\pi}^{2}}{2}\bar{L}^{a}_{t}\bar{L}^{a}_{t}
-\frac{1}{2e^{2}}\bar{L}^{a}_{t}M^{ab}\bar{L}^{b}_{t},  \label{eq:kt}
\end{equation}
where
\begin{equation}
\bar{L}_{\mu}^{a}=R^{a}_{\mu}(\tilde{U}_{1})+L^{a}_{\mu}
(\tilde{U}_{2}),
\end{equation}
and
\begin{equation}
M^{ab}=\bar{L}_{i}^{a}\bar{L}_{i}^{b}-\delta^{ab}
\bar{L}_{i}^{c}\bar{L}_{i}^{c}.
\end{equation}
If we use the hedgehog form for $U_{1}$ and $U_{2}$, we find
\begin{eqnarray}
&& R^{a}_{t}(\tilde{U}_{1})= R^{a}_{i}(\tilde{U}_{1}) T_{1i},
\nonumber \\ &&
 L^{a}_{t}(\tilde{U}_{2})= L^{a}_{i}(\tilde{U}_{2}) T_{2i}, \label{eq:hg1}
\end{eqnarray}
where
\begin{eqnarray}
&&T_{1i}=-v_{1i}+2\epsilon_{ilm}x_{1l}R^{m}_{t}(A_{1}),
\nonumber \\ &&
T_{2i}=-v_{2i}+2\epsilon_{ilm}x_{2l}R^{m}_{t}(A_{2}).    \label{eq:hg2}
\end{eqnarray}
and $\bbox{v}_{1}={d\bbox{x}_{1}}/{dt}=-\frac{1}{2}\dot{\bbox{R}}$ and
$\bbox{v}_{2}={d\bbox{x}_{2}}/{dt}=\frac{1}{2}\dot{\bbox{R}}$. In the
following we will use $\bbox{V}=\dot{\bbox{R}}$ to represent the
relative translational
velocity. The quantities
$\bbox{R}_{t}(A_1)$ and $\bbox{R}_{t}(A_2)$ are iso-rotational
velocities expressed in terms of the right Sugawara variables of the
iso-rotation matrices. For isolated Skyrmions, they are related
to the spin operators of the individual Skyrmions upon quantization by
the relation
\begin{equation}
\bbox{R}_{t}(A_{i})=\frac{\bbox{S}_{i}}{2\Lambda}, \hspace{4em} i=1,2
\end{equation}
where $\Lambda$ is the momenta of inertia for a single Skyrmion.
Since the iso-rotation matrices $A_{1}$ and
$A_{2}$ are global rotations, we have for the spatial components of
the single Skyrmion Sugawara variables
\begin{eqnarray}
&& R^{a}_{i}(\tilde{U}_{1})=D^{ab}(A_{1})R^{b}_{i}(U_{1}),
\nonumber \\ &&
 L^{a}_{i}(\tilde{U}_{2})=D^{ab}(A_{2})L^{b}_{i}(U_{2}),  \label{eq:hg3}
\end{eqnarray}
where $D^{ab}(A_{1})$ and $D^{ab}(A_{2})$ are the Wigner
$D$ functions for $A_{1}$ and $A_{2}$ and in terms of the
components of the $SU(2)$ matrix $A(=A_{1}\mbox{or}A_{2})=a_{4}+i\bbox{\tau}
\cdot\bbox{a}$, they are given by
\begin{equation}
D^{ab}(A)=(a_{4}^2-\bbox{a}^2)\delta_{ab}+2a_{a}a_{b}+2\epsilon_{abc}
a_{c}a_{4}.
\end{equation}
It is easy to verify that for single hedgehog Skyrmions $U_{1}$ and $U_{2}$,
we have
\begin{eqnarray}
&&R^{a}_{i}(U_{1})=x_{1i}x_{1a}P_{1}+\delta_{ia}Q_{1}
-\epsilon_{aic}x_{1c}S_{1},
\nonumber \\ &&
L^{a}_{i}(U_{2})=x_{2i}x_{2a}P_{2}+\delta_{ia}Q_{2}
+\epsilon_{aic}x_{2c}S_{2},
\label{eq:hg4}
\end{eqnarray}
where $P$'s, $Q$'s and $S$'s are scalar functions given by
\begin{eqnarray}
&& P=\frac{1}{x^{2}}(F^{'}-\frac{\sin F\cos F}{x}),
\nonumber \\ &&
Q=\frac{\sin F\cos F}{x},
\nonumber \\ &&
S=\frac{\sin^{2}F}{x^{2}},
\end{eqnarray}
in terms of the hedgehog profile function $F$ as defined in
(\ref{eq:profile}) and its derivative $F^{'}$.

Substituting Eqs.(\ref{eq:hg1}-\ref{eq:hg4}) into (\ref{eq:kt}),
we obtain for  ${\cal K}$
a sum of terms that are quadratic in the velocity vectors
$\bbox{V},\bbox{R}_{t}(A_1)$ and $\bbox{R}_{t}(A_2)$ with coefficients
that depend on $\bbox{x}_{1},\bbox{x}_{2}, \bbox{R}$ and the relative
iso-rotation matrix $C=A_{1}^{\dagger}A_{2}$. The kinetic energy is given by
the integral over the kinetic energy density
\begin{equation}
K=\int d^{3}r {\cal K}.
\end{equation}
The three dimensional integral over all the terms in the expansion of
${\cal K}$ can be reduced to a two dimensional integral
by virtue of the axial symmetry around the axis connecting the
centers of the two Skyrmions. The integrands of these three dimensional
integrals are all of the form $r_{a_{1}}r_{a_{2}}...r_{a_{n}}f(x_{1},x_{2})$
where each $r_{a_{i}}$ is a component of the vector $\bbox{r}$ and $n$
is a non-negative integer.
$f(x_{1},x_{2})$ depends
on $r, R$ and the angle $\theta$ between $\bbox{r}$ and $\bbox{R}$.
It is easy to see that
\begin{equation}
\int d^{3}r\,  r_{a}f(x_{1},x_{2})=\hat{R}_{a}\int (drd\theta) r\cos\theta
f(x_{1},x_{2}),
\end{equation}
where we use the notation
\begin{equation}
\int (drd\theta)=2\pi\int_{0}^{\infty}r^2 dr\int_{0}^{\pi}d\theta\sin\theta.
\end{equation}
It can also be shown that
\begin{eqnarray}
&& \int d^{3}r\,  r_{a}r_{b}f(x_{1},x_{2})=\nonumber\\&&
\hat{R}_{a}\hat{R}_{b}\int (drd\theta) r^2(\cos^2\theta-\frac{1}{2}
\sin^2\theta)f(x_{1},x_{2}) \nonumber \\ &&
+\delta_{ab}\int (drd\theta)r^2 \frac{1}{2}\sin^2\theta f(x_{1},x_{2}).
\end{eqnarray}
and similar relations
for \linebreak[1] integrals \linebreak[2] over \linebreak[3]
$r_{a_{1}}r_{a_{2}}...r_{a_{n}}f(x_{1},x_{2})$ with $n>2$.
We use these relations to write the kinetic energy as a sum of scalar
functions of the velocities, the unit vector $\hat{R}$ of
the relative separation of the two Skyrmions and of the vector that
specifies their relative iso-orientation.

\section{Algebraic Results}
\label{sec:3}

There have been a number of papers dealing with various selected terms in
the kinetic energy. Oka and Odawara {\em et al} \cite{Oka,Yazaki}
have  studied the inertial parameters for translational motion only by
neglecting the time-dependence of iso-rotations of the two
Skyrmions. Riska and his co-workers \cite{Riska}
have investigated terms in
the kinetic energy that couple the translational velocity
with the isorotational velocities and
that contribute to the spin-orbit interaction
when projected  onto nucleon-nucleon states. A similar
study was done by Otofuji {\em et al} \cite{Otofuji}.
However, our recent work \cite{SpinOrbit} has shown that
a proper treatment of the spin-orbit interaction requires inversion of the
mass matrix before projecting it onto nucleon states.  This can not be
done when  terms that couple the translational velocity
with the isorotational velocities are studied alone.
In this work, we present a complete analysis of
the kinetic energy within the product ansatz
by expanding it in a set of rotational
invariant quadratic functions of the translational and iso-rotational
velocities and evaluating all coefficients in the expansion.

The expansion can be conveniently divided into two groups.
The first group, $K^{(i)}_{1}$,
contains terms involving a
set of  functions defined over the translational and
iso-rotational velocity vectors. The second group, $K^{(i)}_{2}$,
is a linear combination
of  another set of functions defined over the three distinct
doublets of the three velocity vectors. In both groups, we order terms
by their irreducible representations under
the relative Skyrmion-Skyrmion iso-orientation,
$C$. The terms quadratic in $\bbox{c}$ and $c_4$ we call $T$ (for two),
and those quartic  we call $F$ (for four).
The $T$'s are the set of four
dimensional harmonic polynomials of degree two and the $F$'s of degree
four.  They correspond to the symmetric representations of
$O(4)$, $(2,0)$ and $(4,0)$, respectively.  Since they are the symmetric
representation, they have spin equal to iso-spin ($S=I$), with
$T$ corresponding to $S=I=1$ and $F$ to $S=I=2$.  The symmetric
representations of $O(4)$,
$(\sigma,0)$, may be further distinguished by their  grand-spin or
$K$-spin, $\bbox{K} = \bbox{S} + \bbox{I}$, where $K$ takes on the
values $0,1,2...,\sigma$.  This then gives for $T$
a scalar term, $T^{(0)}$, a vector term, $T^{(1)}$, and a
rank two term
$T^{(2)}$.
For the forms quartic in $C$,
there are $F^{(0)}$, $F^{(1)}$, $F^{(2)}$, $F^{(3)}$,
and $F^{(4)}$.
Matrix elements of operators
of the $F$-type  vanish between two nucleon states.
The detailed form for $T$ and
$F$ in terms of $C$ are given in the Appendix. Using these expresssions
we arrive at the
following compact form for the expansion of the kinetic energy:
\begin{equation}
K=\sum_{i=0,1,2}(K^{(i)}_{1}+K^{(i)}_{2}), \label{eq:Kexp}
\end{equation}
where
\begin{eqnarray}
K^{(i)}_{1}&=& d^{(i)}_{1}  (\bbox{t}\cdot\bbox{t})
+d^{(i)}_{2} (\bbox{t}\cdot\hat{R})^{2}
+s^{(i)}_{1} (\bbox{t}\cdot\bbox{t}) T^{(0)}
+s^{(i)}_{2} (\bbox{t}\cdot\hat{R})^{2} T^{(0)}
+s^{(i)}_{3} (\bbox{t}\cdot\bbox{t}) T^{(2)}(\hat{R},\hat{R})
\nonumber \\ &&
+s^{(i)}_{4} (\bbox{t}\cdot\hat{R})^{2}  T^{(2)}(\hat{R},\hat{R})
+s^{(i)}_{5}  (\bbox{t}\cdot\hat{R}) T^{(2)}(\hat{R},\bbox{t}) +
s^{(i)}_{6}(\bbox{t}\cdot\hat{R})T^{(2)}
(\hat{R},\hat{R}\times\bbox{t})
\nonumber \\ &&
+s^{(i)}_{7} T^{(2)}(\bbox{t},\bbox{t})
+s^{(i)}_{8}T^{(2)}(\bbox{t},\hat{R}\times\bbox{t})
\nonumber\\&&
+q^{(i)}_{1} (\bbox{t}\cdot\bbox{t}) F^{(0)}
+q^{(i)}_{2} (\bbox{t}\cdot\hat{R})^{2} F^{(0)}
+q^{(i)}_{3} (\bbox{t}\cdot\bbox{t}) F^{(2)}(\hat{R},\hat{R}) +
q^{(i)}_{4} (\bbox{t}\cdot\hat{R})^{2}  F^{(2)}(\hat{R},\hat{R})
\nonumber \\ &&+
q^{(i)}_{5}  (\bbox{t}\cdot\hat{R}) F^{(2)}(\hat{R},\bbox{t})
+q^{(i)}_{6}  (\bbox{t}\cdot\hat{R})
F^{(2)}(\hat{R},\hat{R}\times\bbox{t})
+q^{(i)}_{7} F^{(2)}(\bbox{t},\bbox{t})\nonumber \\ &&
+q^{(i)}_{8}F^{(2)}(\bbox{t},\hat{R}\times\bbox{t})
+q^{(i)}_{9}  (\bbox{t}\cdot\bbox{t}) F^{(4)}(\hat{R},\hat{R},\hat{R},\hat{R})
+q^{(i)}_{10} (\bbox{t}\cdot\hat{R})^{2}
F^{(4)}(\hat{R},\hat{R},\hat{R},\hat{R})
\nonumber \\ &&
+q^{(i)}_{11}  (\bbox{t}\cdot\hat{R}) F^{(4)}(\hat{R},\hat{R},\hat{R},\bbox{t})
+q^{(i)}_{12} (\bbox{t}\cdot\hat{R}) F^{(4)}(\hat{R},\hat{R},\hat{R},
\hat{R}\times\bbox{t})
\nonumber \\ &&
+q^{(i)}_{13} F^{(4)}(\hat{R},\hat{R},\bbox{t},\bbox{t})
+q^{(i)}_{14} F^{(4)}(\hat{R},\hat{R},\bbox{t},\hat{R}\times\bbox{t})
\nonumber \\ &&
+\gamma^{(i)}_{1} (\bbox{t}\cdot\bbox{t})T^{(1)}(\hat{R})
+\gamma^{(i)}_{2} (\bbox{t}\cdot\hat{R})^{2}T^{(1)}(\hat{R})
+\gamma^{(i)}_{3} (\bbox{t}\cdot\hat{R})T^{(1)}(\bbox{t})
\nonumber \\ &&
+\gamma^{(i)}_{4}  (\bbox{t}\cdot\hat{R})T^{(1)}(\hat{R}\times\bbox{t})
\nonumber \\ &&
+ \xi^{(i)}_{1} (\bbox{t}\cdot\bbox{t})F^{(1)}(\hat{R})
+\xi^{(i)}_{2} (\bbox{t}\cdot\hat{R})^{2}F^{(1)}(\hat{R})
+\xi^{(i)}_{3} (\bbox{t}\cdot\hat{R})F^{(1)}(\bbox{t})
\nonumber \\ &&
+\xi^{(i)}_{4}  (\bbox{t}\cdot\hat{R})F^{(1)}(\hat{R}\times\bbox{t})
+\xi^{(i)}_{5}  (\bbox{t}\cdot\bbox{t}) F^{(3)}(\hat{R},\hat{R},\hat{R})
+\xi^{(i)}_{6} (\bbox{t}\cdot\hat{R})^{2} F^{(3)}(\hat{R},\hat{R},\hat{R})
\nonumber \\ &&
+\xi^{(i)}_{7} (\bbox{t}\cdot\hat{R})F^{(3)}(\hat{R},\hat{R},\bbox{t})
+\xi^{(i)}_{8}
(\bbox{t}\cdot\hat{R})F^{(3)}(\hat{R},\hat{R},\hat{R}\times\bbox{t})
\nonumber \\ &&
+\xi^{(i)}_{9} F^{(3)}(\bbox{t},\bbox{t},\hat{R})
+\xi^{(i)}_{10} F^{(3)}(\hat{R},\bbox{t},\hat{R}\times\bbox{t}),
\end{eqnarray}
and
\begin{eqnarray}
K^{(i)}_{2}&=& u^{(i)}_{1}(\bbox{t}_{1}\cdot\bbox{t}_{2})
+u^{(i)}_{2} (\bbox{t}_{1}\cdot\hat{R}) (\bbox{t}_{2}\cdot\hat{R})
+u^{(i)}_{3} [\hat{R}\cdot(\bbox{t}_{1}\times\bbox{t}_{2})]
\nonumber \\ &&
+ v^{(i)}_{1}(\bbox{t}_{1}\cdot\bbox{t}_{2}) T^{(0)}
+v^{(i)}_{2} (\bbox{t}_{1}\cdot\hat{R}) (\bbox{t}_{2}\cdot\hat{R}) T^{(0)}
+v^{(i)}_{3} [\hat{R}\cdot(\bbox{t}_{1}\times\bbox{t}_{2})] T^{(0)}
\nonumber \\ &&
+v^{(i)}_{4}(\bbox{t}_{1}\cdot\bbox{t}_{2}) T^{(2)}(\hat{R},\hat{R})
+v^{(i)}_{5} (\bbox{t}_{1}\cdot\hat{R}) (\bbox{t}_{2}\cdot\hat{R})
T^{(2)}(\hat{R},\hat{R})
\nonumber \\ &&
+v^{(i)}_{6} [\hat{R}\cdot(\bbox{t}_{1}\times\bbox{t}_{2})]
T^{(2)}(\hat{R},\hat{R})
+v^{(i)}_{7}
(\bbox{t}_{1}\cdot\hat{R})T^{(2)}(\hat{R},\hat{R}\times\bbox{t}_{2})
\nonumber \\ &&
+v^{(i)}_{8}
(\bbox{t}_{2}\cdot\hat{R})T^{(2)}(\hat{R},\hat{R}\times\bbox{t}_{1})
+v^{(i)}_{9} (\bbox{t}_{1}\cdot\hat{R})T^{(2)}(\hat{R},\bbox{t}_{2})
+v^{(i)}_{10} (\bbox{t}_{2}\cdot\hat{R})T^{(2)}(\hat{R},\bbox{t}_{1})
\nonumber \\ &&
+v^{(i)}_{11} T^{(2)}(\bbox{t}_{1},\hat{R}\times\bbox{t}_{2})
+
v^{(i)}_{12} T^{(2)}(\bbox{t}_{2},\hat{R}\times\bbox{t}_{1})+
v^{(i)}_{13} T^{(2)}(\hat{R}\times\bbox{t}_{1},\hat{R}\times\bbox{t}_{2})+
\nonumber \\ &&
+  w^{(i)}_{1}(\bbox{t}_{1}\cdot\bbox{t}_{2}) F^{(0)}
+w^{(i)}_{2} (\bbox{t}_{1}\cdot\hat{R}) (\bbox{t}_{2}\cdot\hat{R}) F^{(0)}
+w^{(i)}_{3} [\hat{R}\cdot(\bbox{t}_{1}\times\bbox{t}_{2})] F^{(0)}
\nonumber \\ &&
+w^{(i)}_{4}(\bbox{t}_{1}\cdot\bbox{t}_{2}) F^{(2)}(\hat{R},\hat{R})
+w^{(i)}_{5} (\bbox{t}_{1}\cdot\hat{R}) (\bbox{t}_{2}\cdot\hat{R})
F^{(2)}(\hat{R},\hat{R})
\nonumber \\ &&
+w^{(i)}_{6} [\hat{R}\cdot(\bbox{t}_{1}\times\bbox{t}_{2})]
F^{(2)}(\hat{R},\hat{R})
\nonumber \\ &&
+w^{(i)}_{7}
(\bbox{t}_{1}\cdot\hat{R})F^{(2)}(\hat{R},\hat{R}\times\bbox{t}_{2})
+w^{(i)}_{8}
(\bbox{t}_{2}\cdot\hat{R})F^{(2)}(\hat{R},\hat{R}\times\bbox{t}_{1})
\nonumber \\ &&
+w^{(i)}_{9} (\bbox{t}_{1}\cdot\hat{R})F^{(2)}(\hat{R},\bbox{t}_{2})
+w^{(i)}_{10} (\bbox{t}_{2}\cdot\hat{R})F^{(2)}(\hat{R},\bbox{t}_{1})
+w^{(i)}_{11} F^{(2)}(\bbox{t}_{1},\hat{R}\times\bbox{t}_{2})\nonumber \\ &&
+
w^{(i)}_{12} F^{(2)}(\bbox{t}_{2},\hat{R}\times\bbox{t}_{1})+
w^{(i)}_{13} F^{(2)}(\hat{R}\times\bbox{t}_{1},\hat{R}\times\bbox{t}_{2})+
w^{(i)}_{14} (\bbox{t}_{1}\cdot\bbox{t}_{2})
F^{(4)}(\hat{R},\hat{R},\hat{R},\hat{R})
\nonumber \\ &&
+w^{(i)}_{15}
 (\bbox{t}_{1}\cdot\hat{R}) (\bbox{t}_{2}\cdot\hat{R})
F^{(4)}(\hat{R},\hat{R},\hat{R},\hat{R})
+w^{(i)}_{16} [\hat{R}\cdot(\bbox{t}_{1}\times\bbox{t}_{2})]
F^{(4)}(\hat{R},\hat{R},\hat{R},\hat{R})
\nonumber \\ &&
+w^{(i)}_{17} (\bbox{t}_{1}\cdot\hat{R}) F^{(4)}(\hat{R},\hat{R},\hat{R},
\bbox{t}_{2})
+w^{(i)}_{18}  (\bbox{t}_{2}\cdot\hat{R}) F^{(4)}(\hat{R},\hat{R},\hat{R},
\bbox{t}_{1})\nonumber \\ &&
+ w^{(i)}_{19}
(\bbox{t}_{1}\cdot\hat{R}) F^{(4)}(\hat{R},\hat{R},\hat{R},\hat{R}\times
\bbox{t}_{2})
+w^{(i)}_{20}(\bbox{t}_{2}\cdot\hat{R})
F^{(4)}(\hat{R},\hat{R},\hat{R},\hat{R}\times\bbox{t}_{1})
\nonumber \\ &&
+w^{(i)}_{21}F^{(4)}( \hat{R},\hat{R},\bbox{t}_{1},\hat{R}\times
\bbox{t}_{2})
+w^{(i)}_{22}F^{(4)}( \hat{R},\hat{R},\bbox{t}_{2},\hat{R}\times
\bbox{t}_{1}) \nonumber \\ &&
+w^{(i)}_{23}F^{(4)}(\hat{R},\hat{R},\hat{R}\times\bbox{t}_{1},
\hat{R}\times\bbox{t}_{2})
\nonumber \\ &&
+\phi^{(i)}_{1} (\bbox{t}_{1}\cdot\bbox{t}_{2}) T^{(1)}(\hat{R})
+\phi^{(i)}_{2}  (\bbox{t}_{1}\cdot\hat{R}) (\bbox{t}_{2}\cdot\hat{R})
T^{(1)}(\hat{R})
+\phi^{(i)}_{3}  [\hat{R}\cdot(\bbox{t}_{1}\times\bbox{t}_{2})]T^{(1)}(\hat{R})
\nonumber \\ &&
+\phi^{(i)}_{4} (\bbox{t}_{1}\cdot\hat{R}) T^{(1)}(\bbox{t}_{2})
+\phi^{(i)}_{5} (\bbox{t}_{2}\cdot\hat{R}) T^{(1)}(\bbox{t}_{1})
\nonumber \\ &&
+\phi^{(i)}_{6} (\bbox{t}_{1}\cdot\hat{R}) T^{(1)}(\hat{R}\times\bbox{t}_{2})
+\phi^{(i)}_{7} (\bbox{t}_{2}\cdot\hat{R}) T^{(1)}(\hat{R}\times\bbox{t}_{1})
\nonumber \\ &&
+\eta^{(i)}_{1} (\bbox{t}_{1}\cdot\bbox{t}_{2}) F^{(1)}(\hat{R})
+\eta^{(i)}_{2}  (\bbox{t}_{1}\cdot\hat{R}) (\bbox{t}_{2}\cdot\hat{R})
F^{(1)}(\hat{R})
+\eta^{(i)}_{3}  [\hat{R}\cdot(\bbox{t}_{1}\times\bbox{t}_{2})]F^{(1)}(\hat{R})
\nonumber \\ &&
+\eta^{(i)}_{4} (\bbox{t}_{1}\cdot\hat{R}) F^{(1)}(\bbox{t}_{2})
+\eta^{(i)}_{5} (\bbox{t}_{2}\cdot\hat{R}) F^{(1)}(\bbox{t}_{1})
+\eta^{(i)}_{6} (\bbox{t}_{1}\cdot\hat{R}) F^{(1)}(\hat{R}\times\bbox{t}_{2})
\nonumber \\ &&
+\eta^{(i)}_{7} (\bbox{t}_{2}\cdot\hat{R}) F^{(1)}(\hat{R}\times\bbox{t}_{1})
+\eta^{(i)}_{8} (\bbox{t}_{1}\cdot\bbox{t}_{2})
F^{(3)}(\hat{R},\hat{R},\hat{R})
\nonumber \\ &&
+\eta^{(i)}_{9}  (\bbox{t}_{1}\cdot\hat{R})
(\bbox{t}_{2}\cdot\hat{R}) F^{(3)}(\hat{R},\hat{R},\hat{R})
+\eta^{(i)}_{10}  [\hat{R}\cdot(\bbox{t}_{1}\times\bbox{t}_{2})]
 F^{(3)}(\hat{R},\hat{R},\hat{R})
\nonumber \\ &&
+\eta^{(i)}_{11} (\bbox{t}_{1}\cdot\hat{R})
F^{(3)}(\hat{R},\hat{R},\bbox{t}_{2})
+\eta^{(i)}_{12} (\bbox{t}_{2}\cdot\hat{R})
F^{(3)}(\hat{R},\hat{R},\bbox{t}_{1})
\nonumber \\ &&
+\eta^{(i)}_{13} (\bbox{t}_{1}\cdot\hat{R})
F^{(3)}(\hat{R},\hat{R},\hat{R}\times\bbox{t}_{2})
+\eta^{(i)}_{14} (\bbox{t}_{2}\cdot\hat{R})
F^{(3)}(\hat{R},\hat{R},\hat{R}\times\bbox{t}_{1})
\nonumber \\ &&
+\eta^{(i)}_{15} F^{(3)}(\hat{R},\bbox{t}_{1},\bbox{t}_{2})
+\eta^{(i)}_{16} F^{(3)}(\hat{R},\bbox{t}_{1},\hat{R}\times\bbox{t}_{2})
\nonumber \\ &&
+\eta^{(i)}_{17} F^{(3)}(\hat{R},\bbox{t}_{2},\hat{R}\times\bbox{t}_{1})
+\eta^{(i)}_{18}
F^{(3)}(\hat{R},\hat{R}\times\bbox{t}_{1},\hat{R}\times\bbox{t}_{2}).
\end{eqnarray}
For $K^{(i)}_{1}$, we take
\begin{equation}
\begin{array}{ll}
\bbox{t}=\bbox{V}       &,~ i=0, \\
\bbox{t}=\bbox{R}_{t}(A_{1})   &,~ i=1, \\
\bbox{t}=\bbox{R}_{t}(A_{2})   &,~ i=2,
\end{array}
\end{equation}
and for $K^{(i)}_{2}$, we take
\begin{equation}
\begin{array}{ll}
(\bbox{t}_{1},\bbox{t}_{2})=(\bbox{R}_{t}(A_{1}), \bbox{R}_{t}(A_{2})) &,~ i=0,
\\
(\bbox{t}_{1},\bbox{t}_{2})=(\bbox{V},\bbox{R}_{t}(A_{1}))   &,~ i=1, \\
(\bbox{t}_{1},\bbox{t}_{2})=(\bbox{V},\bbox{R}_{t}(A_{2}))   &,~ i=2.
\end{array}
\end{equation}

Careful readers may wonder why a term such as $T^{(2)} (\hat{R}\times\bbox{t},
\hat{R}\times\bbox{t})$ is not present in the expansion above.  It is
because this term can be written as a linear combination of other
terms in the expansion. We eliminate this and other terms that are
linearly dependent on the invariant functions used in the expansion.

We have used Roman characters for coefficients of those terms that are
independent of $c_{4}$ and Greek letters for those terms that
involve $c_{4}$. We find that for the pair
$\left\{K_{1}(\bbox{R}_{t}(A_{1})),
K_{1}(\bbox{R}_{t}(A_{2}))\right\}$ and the pair
$\left\{K_{2}(\bbox{V},\bbox{R}_{t}(A_{1})),
K_{2}(\bbox{V},\bbox{R}_{t}(A_{2}))\right\}$
all Roman letter coefficients have the same sign for the
two terms in each pair while all the Greek letter coefficients
have the opposite sign. This implies that terms that
violate parity are also anti-symmetric under the exchange
of the two Skyrmions.

In the expansion above,
we have included parity-violating terms
that are artifacts of the product ansatz. Of course in an exact
calculation, these terms will not appear.

\section{Numerical Results}
\label{sec:4}

In this section we present the $R$ dependence of the leading terms%
\footnote{ We show only the leading terms here, but all terms
have been calculated and they are available to any one who is interested.}
in the expansion of the kinetic energy, $K_1$ and $K_2$, of the last
section. The kinetic energy is a general bilinear form in the
three velocity vectors, the relative translational velocity,
and the angular velocity of each of the individual Skyrmions.
The first of these we call $\bbox{V}$, the angular velocities
we call $\bbox{R}_{t}(A_i)$.
The parameters of the Skyrme model we use are
taken from \cite{ANW} and are listed in Table 1.
\begin{table}
\caption{Skyrme model parameters}
\begin{center}
\begin{tabular}{ccc}
$m_{\pi}$ (MeV)  & $f_{\pi}$ (MeV)   &   $e$    \\ \hline
0        &   64.5            &   5.45   \\
138      &   54.1            &   4.84   \\
\end{tabular}
\end{center}
\end{table}
The coefficients for terms that are quadratic in $\bbox{V}$
are given in units of MeV since $\bbox{V}$ is given in units of c,
the speed of light.
The coefficients for terms that are linear in
$\bbox{V}$ and $\bbox{R}_{t}(A_{i})$ are given in units of $\mbox{MeV}
\cdot\mbox{fm}$ since $\bbox{R}_{t}(A_i)/c$ has the dimension of
1/fm. It follows that  the coefficients for terms that are quadratic in
$\bbox{R}_{t}(A_{i})$ have units of $\mbox{MeV}\cdot\mbox{fm}^2$.

\subsection{Velocities}

We first consider the terms in $K^{(0)}_1$ in which $\bbox{t} = \bbox{V}$, the
relative velocity between the Skyrmions.  We consider the dependence on
$R$ of each of the coefficients of the invariants in (\ref{eq:Kexp}).
\begin{figure}
\epsfysize=5cm
\centerline{\epsffile{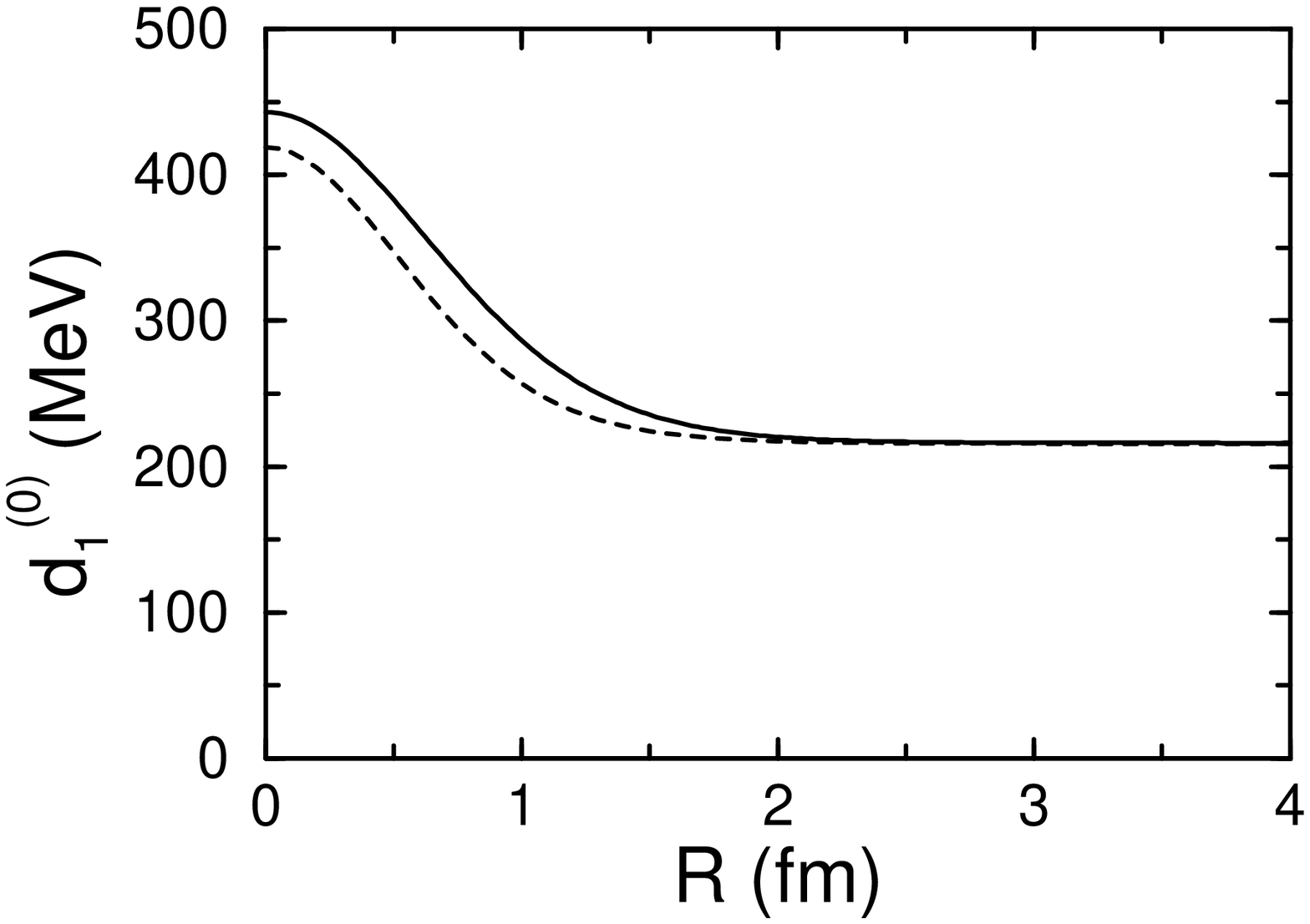}}
\caption{Solid line: $d_{1}^{(0)}$, with pion mass term;
dashed:$d_{1}^{(0)}$, zero pion mass.\label{fig:1}}
\end{figure}
 The first terms, $d^{(0)}_1$ and $d^{(0)}_2$,
do not depend on the relative iso-orientation,
$C$.  We find that $d^{(0)}_1$ is quite large while $d^{(0)}_2$ is small.
The $d^{(0)}_1$
term goes over to a constant at large $R$, since it becomes the ordinary
translational kinetic energy of the single separated Skyrmions.  In that
case, and in our units and conventions, $d^{(0)}_1$ is just $1/4$ the mass
of a single Skyrmion.  We show $d^{(0)}_1$ plotted as a function of $R$ in
Fig.~1. The solid line is the calculation with the physical pion mass
and the dotted line with zero pion mass.  The effect of the pion mass
is relatively small, and only significant inside 1 fm, where we should
distrust the product ansatz.
\begin{figure}
\epsfysize=5cm
\centerline{\epsffile{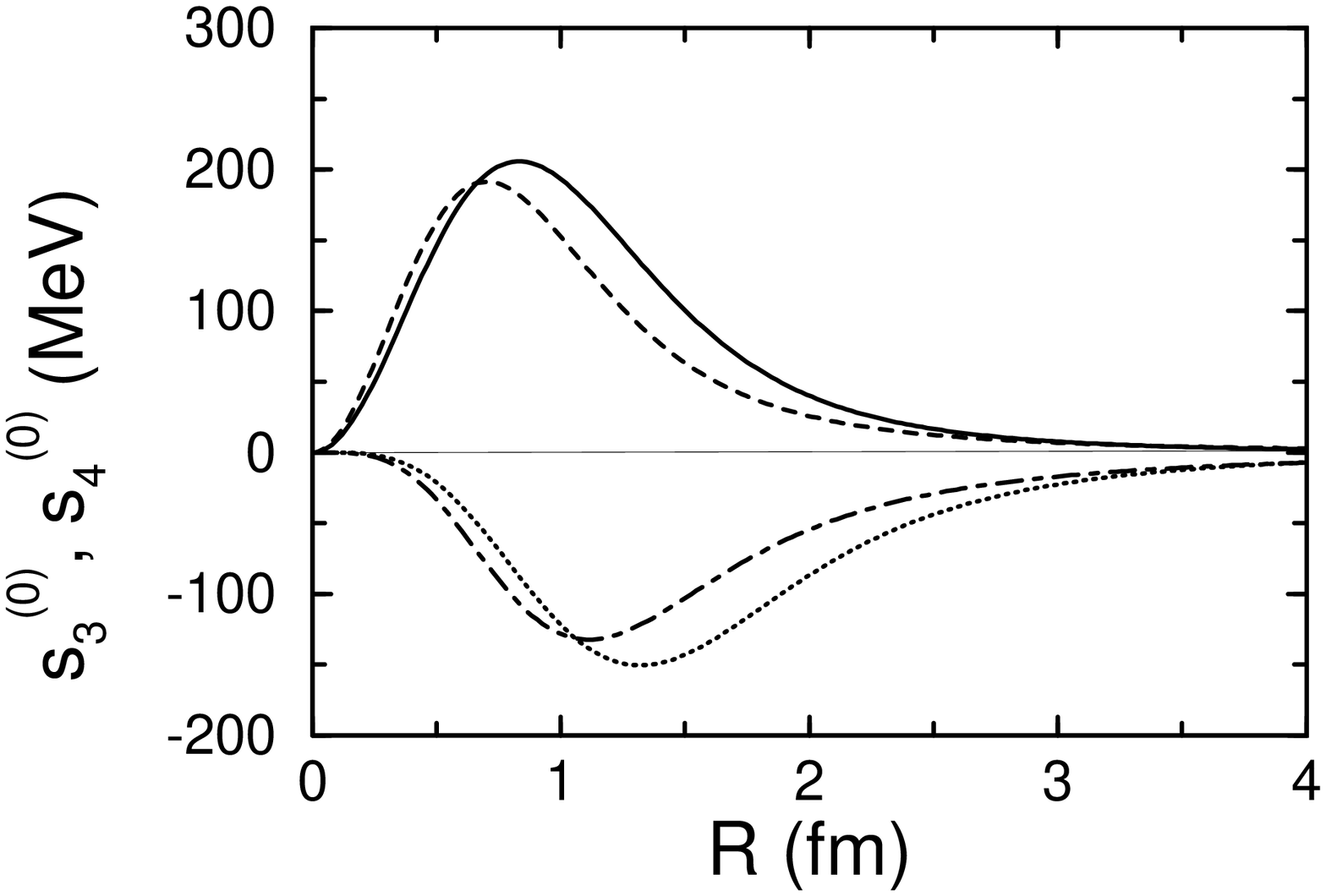}}
\caption{Solid line: $s_{3}^{(0)}$, with pion mass term; dashed line:
$s_{3}^{(0)}$, zero pion mass; dotted: $s_{4}^{(0)}$, with pion mass
term; dot-dashed: $s_{4}^{(0)}$, with zero pion mass.\label{fig:2}}
\end{figure}
\begin{figure}
\epsfysize=5cm
\centerline{\epsffile{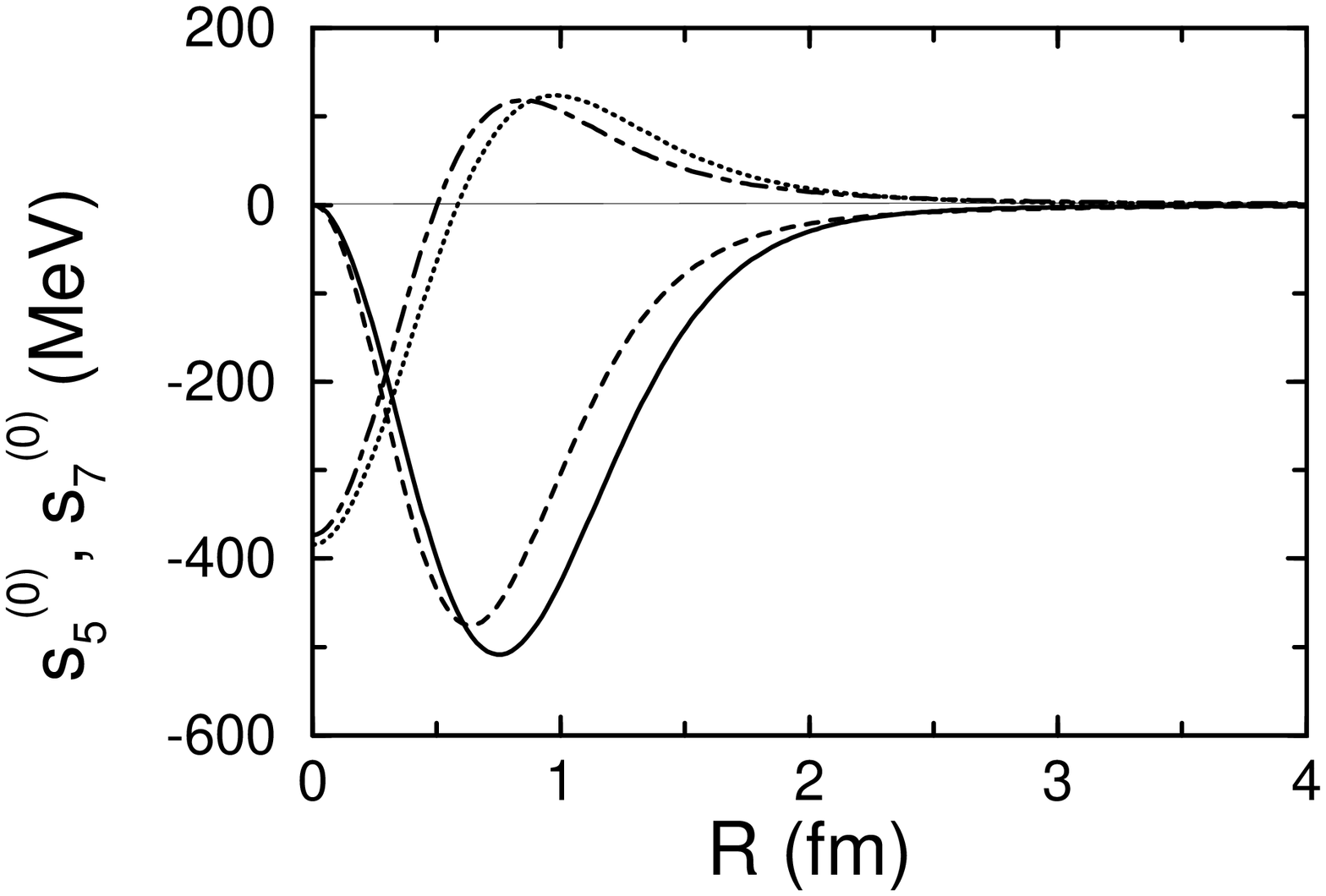}}
\caption{Solid line: $s_{5}^{(0)}$, with pion mass term; dashed line:
$s_{5}^{(0)}$, zero pion mass; dotted: $s_{7}^{(0)}$, with pion mass
term; dot-dashed: $s_{7}^{(0)}$, with zero pion mass.\label{fig:3}}
\end{figure}
\begin{figure}
\epsfysize=5cm
\centerline{\epsffile{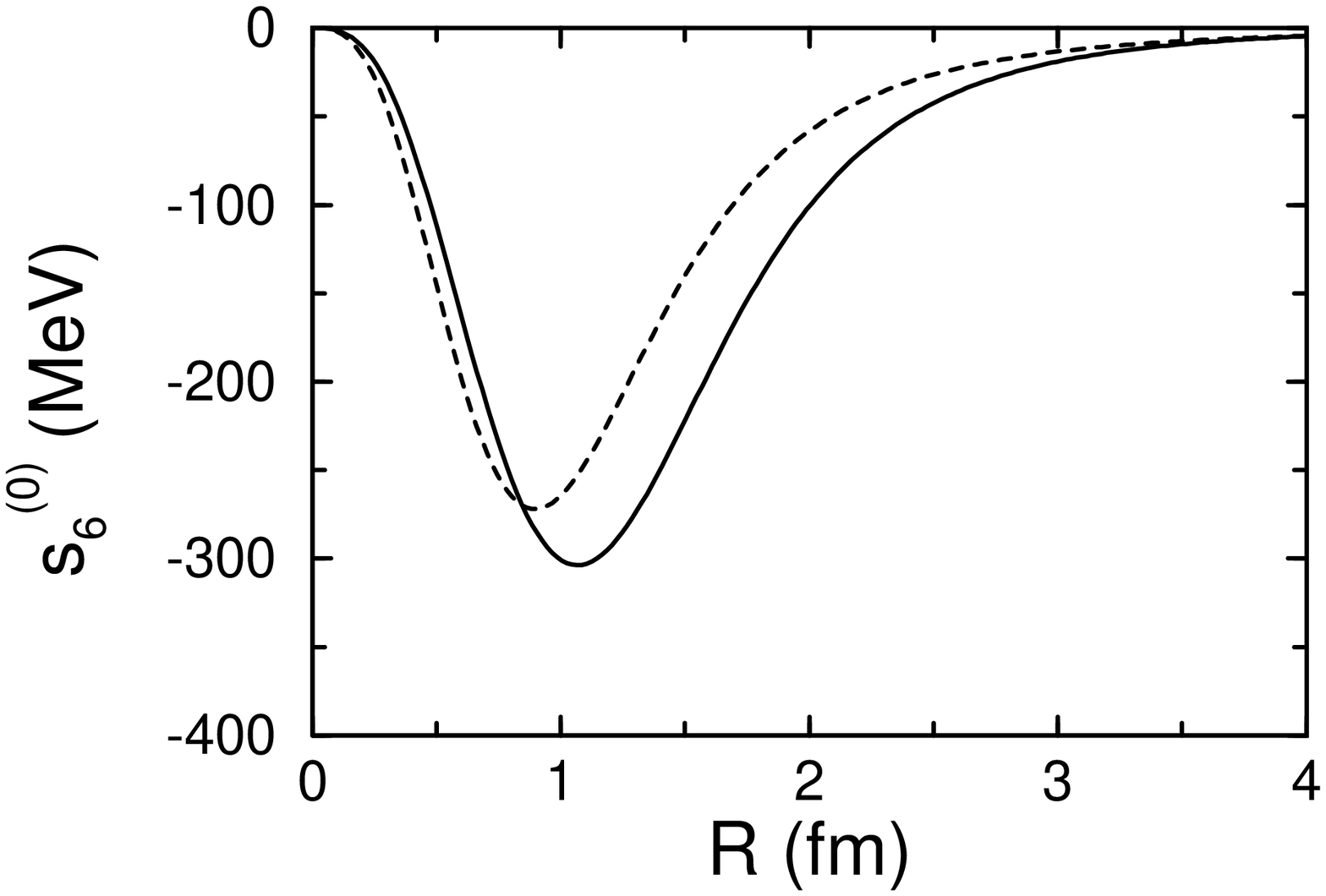}}
\caption{Solid line : $s_{6}^{(0)}$, with pion mass term; dashed line:
$s_{6}^{(0)}$, zero pion mass. This term is odd under parity.
\label{fig:4}}
\end{figure}
We find that $s^{(0)}_1$ and $s^{(0)}_2$ are quite small,
$s^{(0)}_3$ and $s^{(0)}_4$ of moderate size
and $s^{(0)}_5, s^{(0)}_6$ and $s^{(0)}_7$ large at small to
medium values of $R$ (see Figs.~2--4).  $s^{(0)}_6$ is odd under parity and
interchange of the two Skyrmions. No such terms should appear in a
correct calculation, and its large size out to medium $R$ is a measure of
the limitations of the product ansatz.  All further terms in the expansion
of $K^{(0)}_1$ with $\bbox{t} = \bbox{V}$
are quite small outside $R=1$ fm, with
the $q^{(0)}_i$ terms particularly small.
These are $F$ terms (quartic in $C$)
and their small size is a manifestation of the decreasing size of terms with
increasing iso-spin complexity.
The $\gamma^{(0)}_i$ and $\xi^{(0)}$  terms are
all identically zero because $K^{(0)}_{1}$ is in fact invariant under
the transformation $C\rightarrow C^{\dagger}$.

\subsection{Iso-rotation}

Next we consider $\bbox{t}=\bbox{R}_{t}(A_1)$ in $K^{(1)}_1$. The results for
$\bbox{t} = \bbox{R}_{t}(A_2)$ are the same up to terms that are not
symmetric under interchange of the two Skyrmions.
These terms  violate parity.
The term in $d^{(1)}_1$ is a pure iso-kinetic energy.
Since for an individual Skyrmion an iso-rotation is equivalent to a
spin rotation, this term goes over  to the kinetic energy of
a freely rotating Skyrmion for large $R$. There
$d^{(1)}_1$ is equal to twice
the moment of inertia $2\Lambda$.
\begin{figure}
\epsfysize=5cm
\centerline{\epsffile{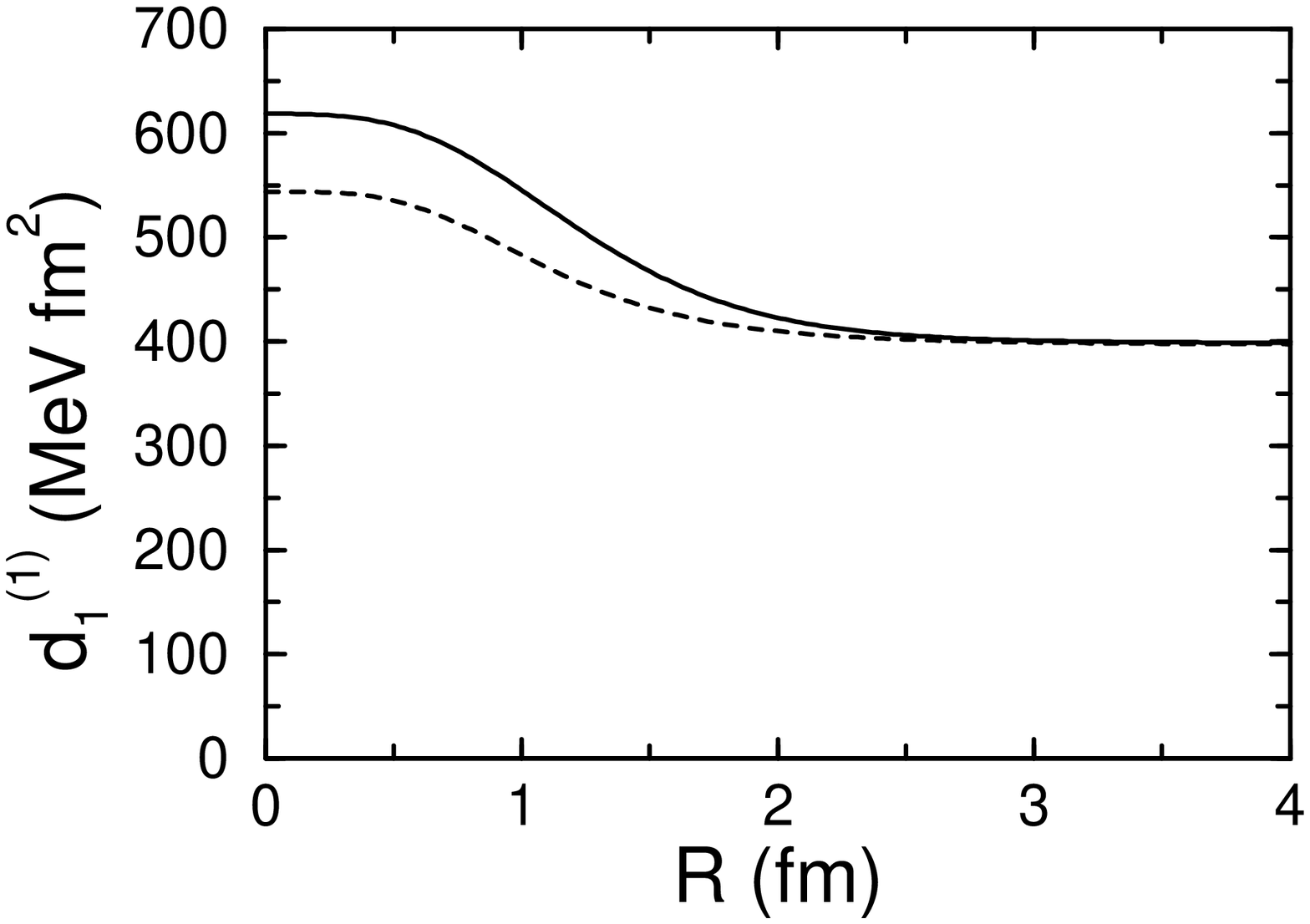}}
\caption{Solid line : $d_{1}^{(1)}$, with pion mass term;
dashed:$d_{1}^{(1)}$, zero pion mass.\label{fig:5}}
\end{figure}
All this is seen in Fig.~5 where we also see that there is more
sensitivity to the pion mass than in the $\bbox{t} = \bbox{V}$ terms, but
mostly at small distances where the product ansatz is not to be trusted.
We find that $d^{(1)}_2$, $s^{(1)}_1$,
and $s^{(1)}_2$ are all quite small except at
very small $R$.
$s^{(1)}_3$, $s^{(1)}_4$, $s^{(1)}_5$, and $s^{(1)}_7$
are all of moderate size beyond
1 fm, and show interesting $R$ dependence  and some sensitivity to the
pion mass (see Figs.~6--7).
\begin{figure}
\epsfysize=5cm
\centerline{\epsffile{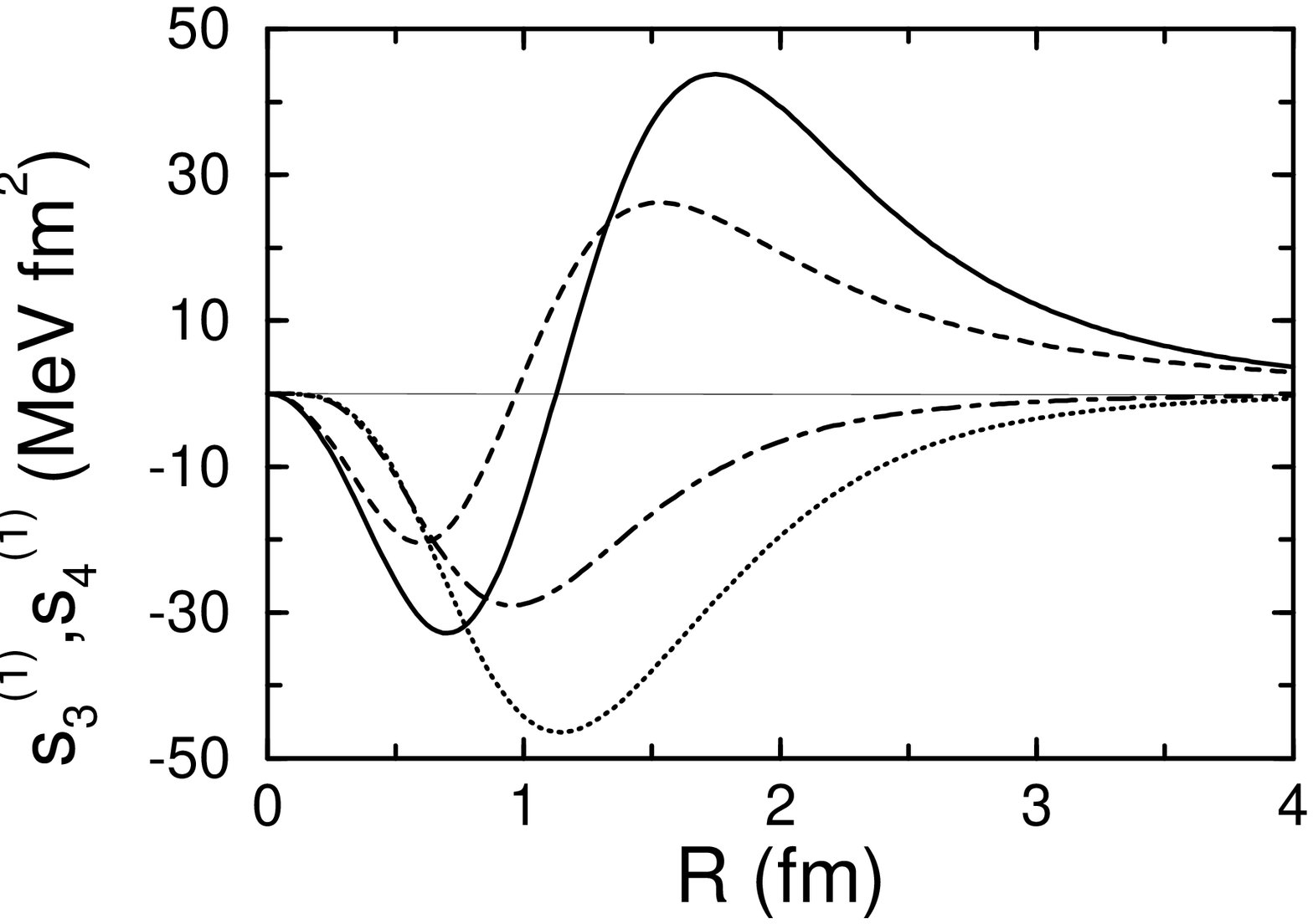}}
\caption{Solid line: $s_{3}^{(1)}$, with pion mass term; dashed line:
$s_{3}^{(1)}$, zero pion mass; dotted: $s_{4}^{(1)}$, with pion mass
term; dot-dashed: $s_{4}^{(1)}$, with zero pion mass.\label{fig:6}}
\end{figure}
\begin{figure}
\epsfysize=5cm
\centerline{\epsffile{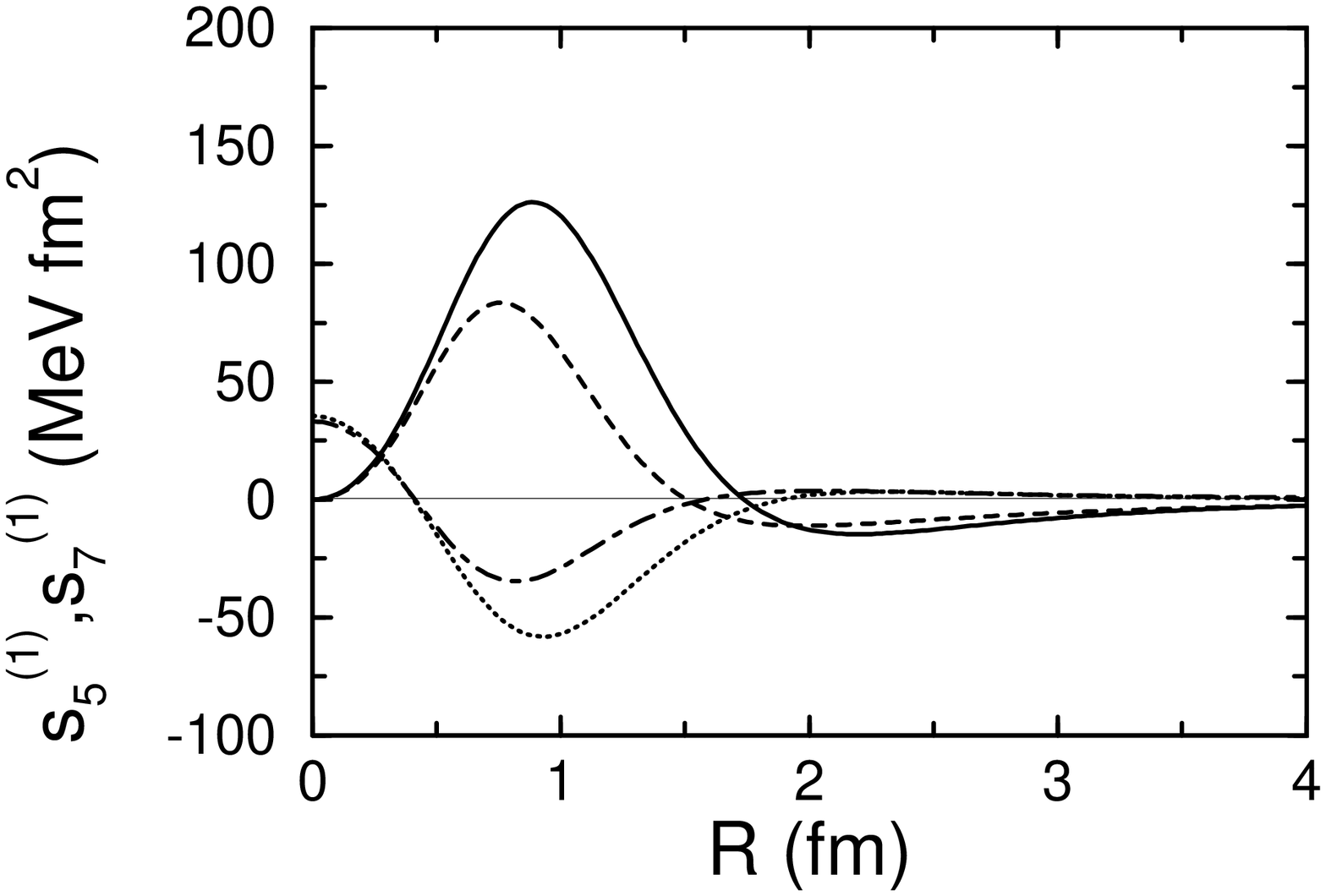}}
\caption{Solid line: $s_{5}^{(1)}$, with pion mass term; dashed line:
$s_{5}^{(1)}$, zero pion mass; dotted: $s_{7}^{(1)}$, with pion mass
term; dot-dashed: $s_{7}^{(1)}$, with zero pion mass.\label{fig:7}}
\end{figure}
\begin{figure}
\epsfysize=5cm
\centerline{\epsffile{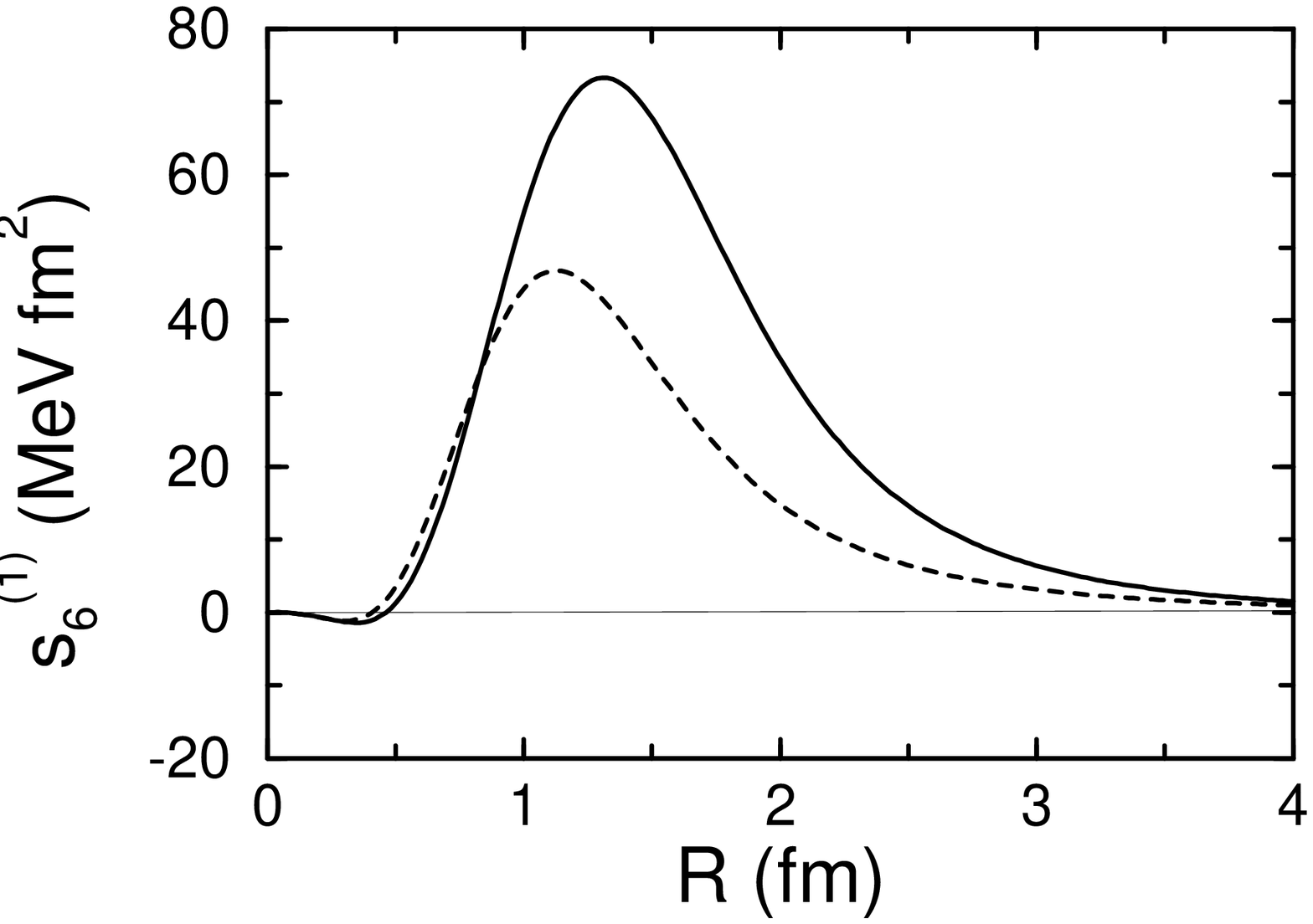}}
\caption{Solid line: $s_{6}^{(1)}$, with pion mass term; dashed line:
$s_{6}^{(1)}$, zero pion mass. This term is odd under parity.
\label{fig:8}}
\end{figure}
$s^{(1)}_6$ (see Fig.~8)
is a wrong parity term that is
comparable with $s^{(1)}_3$, $s^{(1)}_4$, $s^{(1)}_5$, and $s^{(1)}_7$.
The next term that is significant
beyond 1 fm is $q^{(1)}_5$.
It is one of the few quartic terms in $C$ that
is large (see Fig.~9).
\begin{figure}
\epsfysize=5cm
\centerline{\epsffile{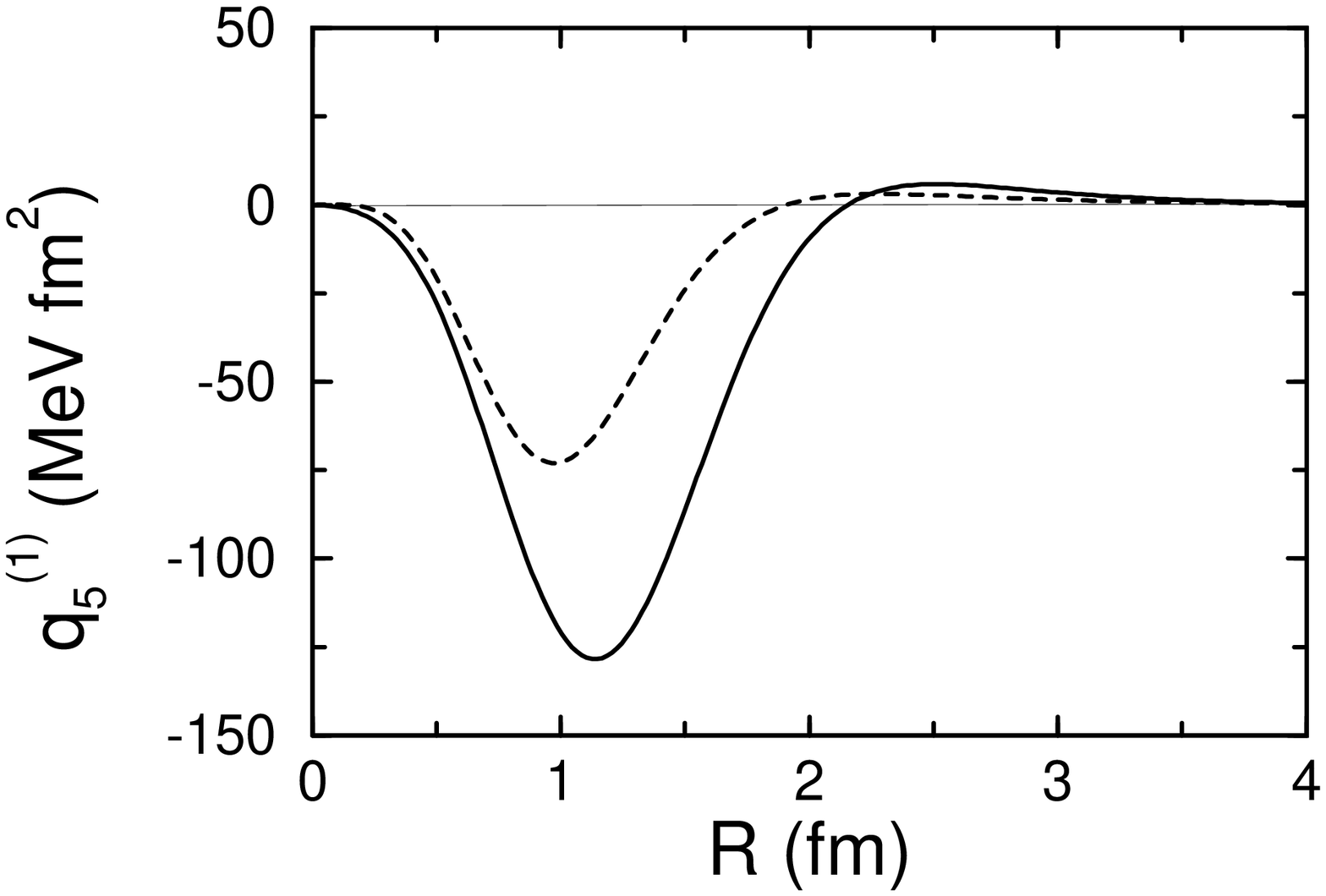}}
\caption{Solid line: $q_{5}^{(1)}$, with pion mass term; dashed line:
$q_{5}^{(1)}$, zero pion mass.\label{fig:9}}
\end{figure}
It is also remarkably sensitive to the inclusion of
a pion mass term.  The
remaining quartic terms are small as are the Greek letter terms.
The largest of these are $\gamma^{(1)}_4$
and $\xi^{(1)}_{10}$, which we show in Fig.~10.
\begin{figure}
\epsfysize=5cm
\centerline{\epsffile{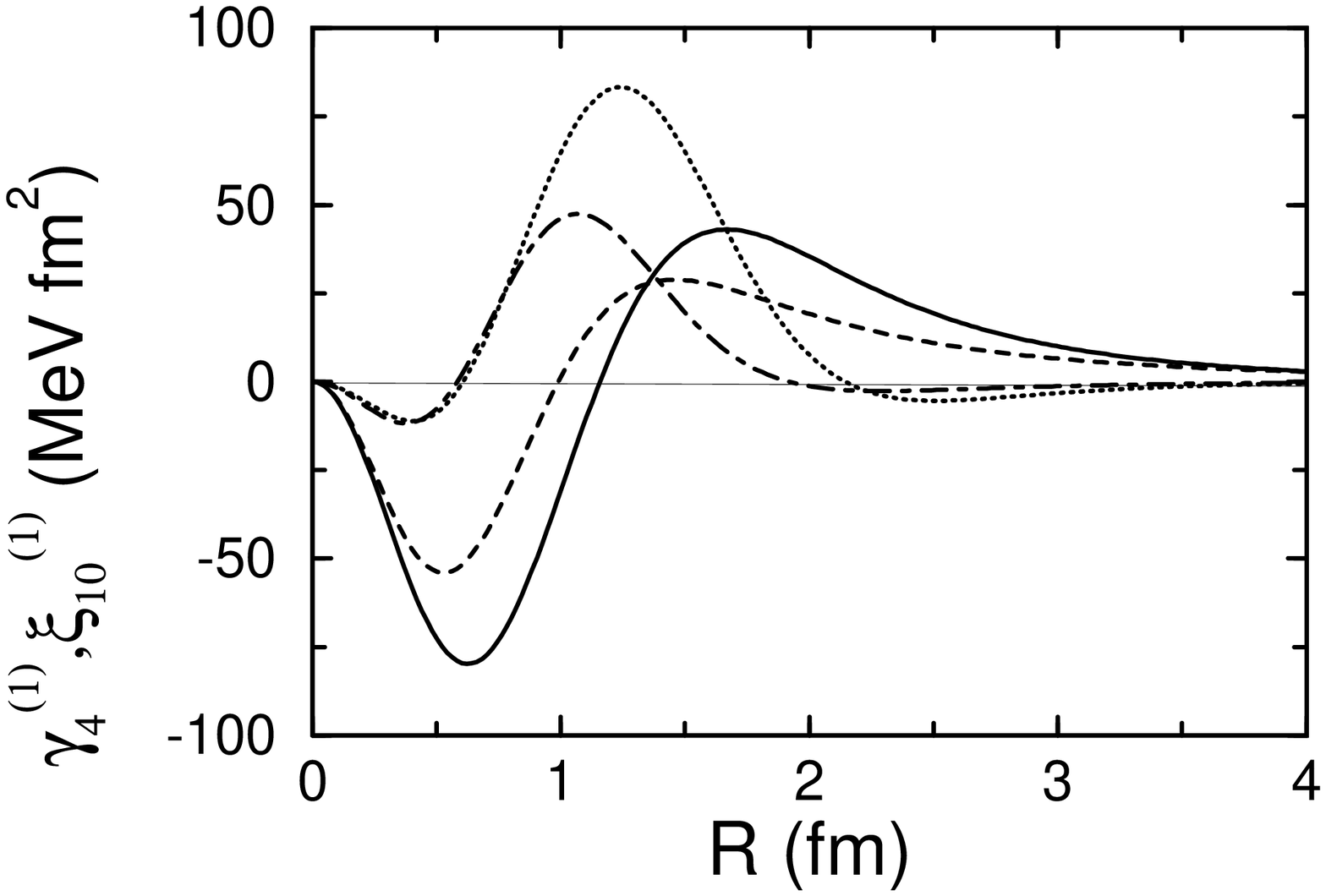}}
\caption{Solid line: $\gamma_{4}^{(1)}$, with pion mass term; dashed line:
$\gamma_{4}^{(1)}$, zero pion mass; dotted: $\xi_{10}^{(1)}$, with pion mass
term; dot-dashed: $\xi_{10}^{(1)}$, with zero pion mass.\label{fig:10}}
\end{figure}

\subsection{Velocity and Iso-rotation Coupling}

We now turn to the terms in $K_2$.  In these terms two kinds of
velocities are coupled.  We consider first $\bbox{t}_{1} = \bbox{V}$ and
$\bbox{t}_{2} = \bbox{R}_{t}(A_1)$. Again the contribution with
$\bbox{t}_{2} = \bbox{R}_{t}(A_2)$ can be obtained from the first by exchanging
one and two with the understanding that terms that violate  parity
are also odd under the interchange.
We find that $u^{(1)}_1$ and $u^{(1)}_2$ are zero.
These terms would violate parity
since $\bbox{V}$ and $\bbox{R}_{t}(A_i)$ behave oppositely under parity.
We find that $u^{(1)}_3$ is relatively large (see Fig.~11).
\begin{figure}
\epsfysize=5cm
\centerline{\epsffile{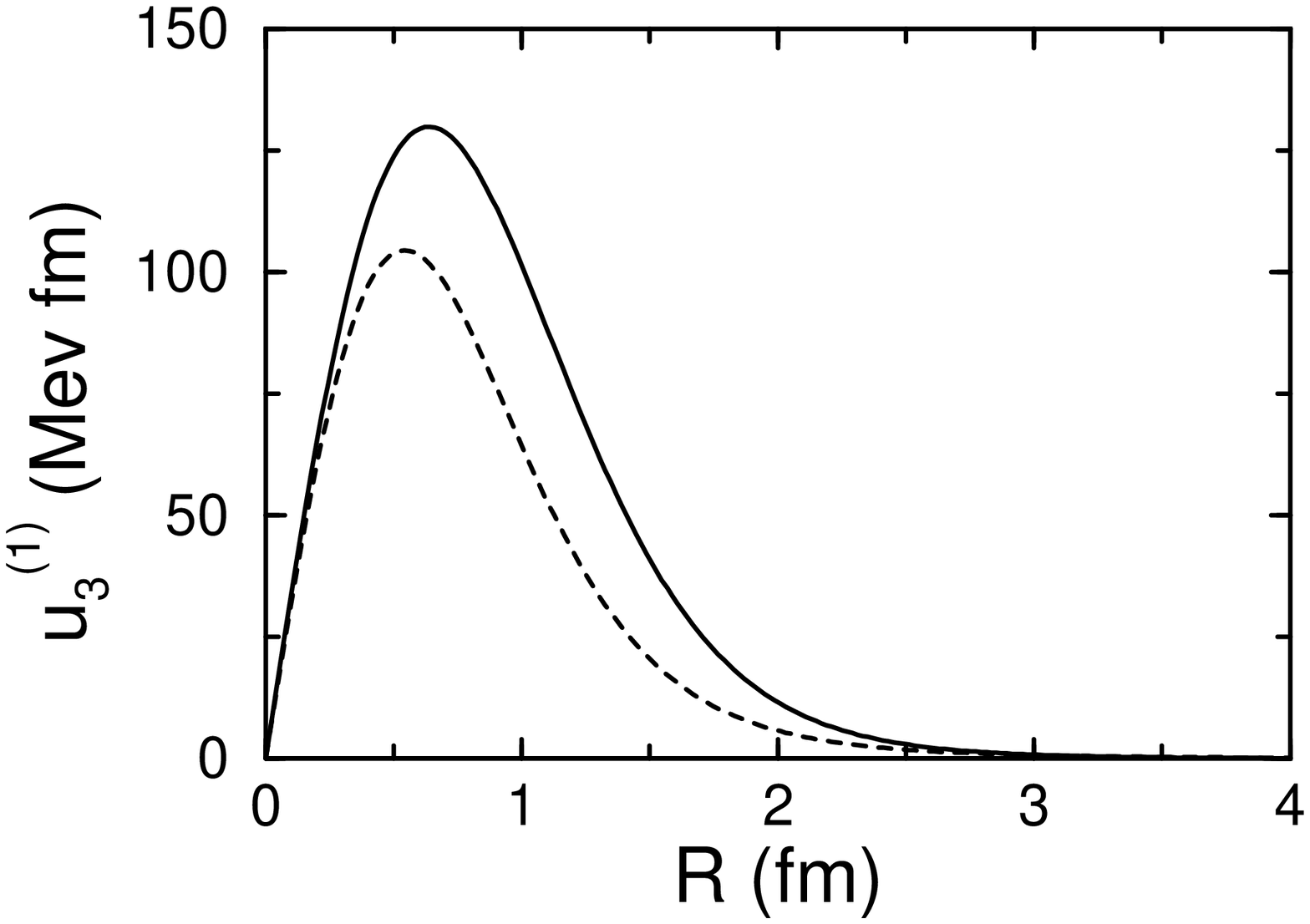}}
\caption{Solid line: $u_{3}^{(1)}$, with pion mass term; dashed line:
$u_{3}^{(1)}$, zero pion mass.\label{fig:11}}
\end{figure}
It is
the term principally responsible for iso-spin independent spin-orbit coupling.
It has been calculated before \cite{Riska,Otofuji} but our result is about
as twice as large as reported there. More importantly even though we obtain the
same sign for  $u^{(1)}_3$ as reported in \cite{Riska,Otofuji},
a careful conversion from
a velocity to a canonical momentum form leads to the correct sign for
the spin-orbit interaction, as we have shown in our recent work
\cite{SpinOrbit}.
$v^{(1)}_1$, $v^{(1)}_2$ and $v^{(1)}_3$
are very small, but $v^{(1)}_4$ and $v^{(1)}_5$ are large at small $R$, and
parity violating  (see Fig.~12).
\begin{figure}
\epsfysize=5cm
\centerline{\epsffile{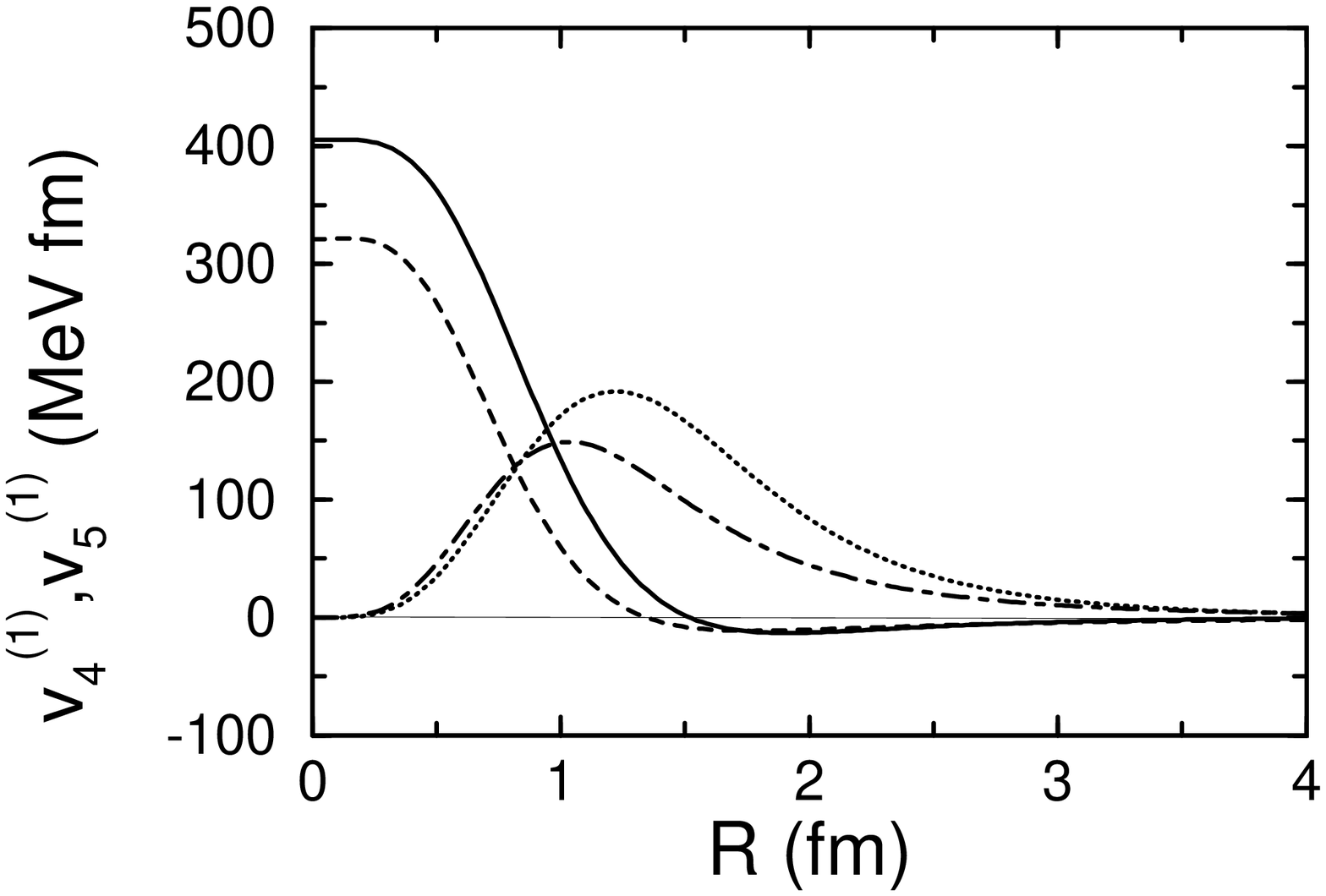}}
\caption{Solid line: $v_{4}^{(1)}$ with pion mass term; dashed line:
$v_{4}^{(1)}$, zero pion mass; dotted: $v_{5}^{(1)}$, with pion mass
term; dot-dashed: $v_{5}^{(1)}$, with zero pion mass.
These terms are odd under parity.\label{fig:12}}
\end{figure}
 $v^{(1)}_6$, $v^{(1)}_7$, $v^{(1)}_8$, $v^{(1)}_{11}$  and $v^{(1)}_{12}$ are
moderately large, but not as large as $u^{(1)}_3$,
and contribute to the iso-spin
dependent spin-orbit interaction (see Figs.~13--15).
\begin{figure}
\epsfysize=5cm
\centerline{\epsffile{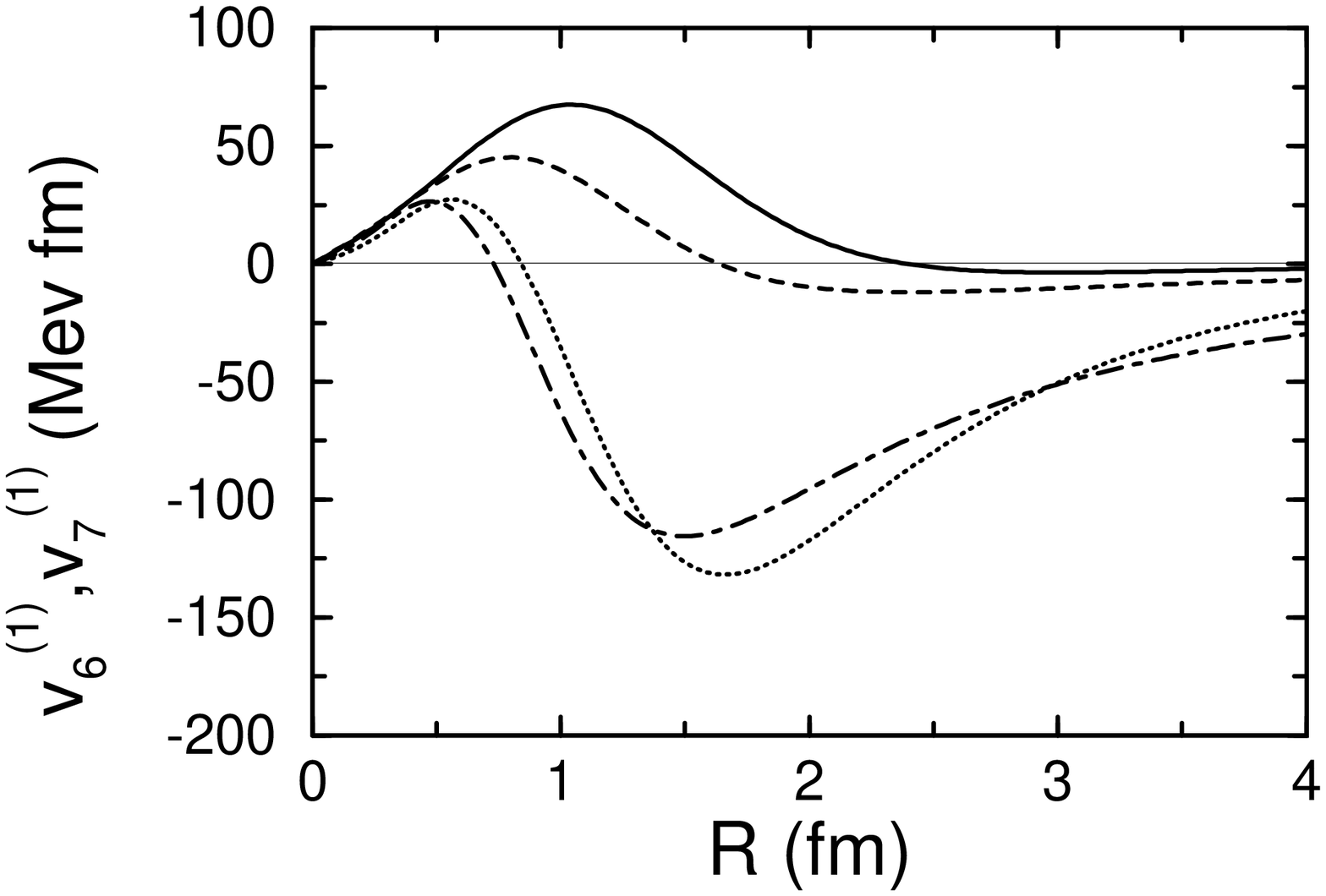}}
\caption{Solid line: $v_{6}^{(1)}$ with pion mass term; dashed line:
$v_{6}^{(1)}$, zero pion mass; dotted: $v_{7}^{(1)}$, with pion mass
term; dot-dashed: $v_{7}^{(1)}$, with zero pion mass.\label{fig:13}}
\end{figure}
\begin{figure}
\epsfysize=5cm
\centerline{\epsffile{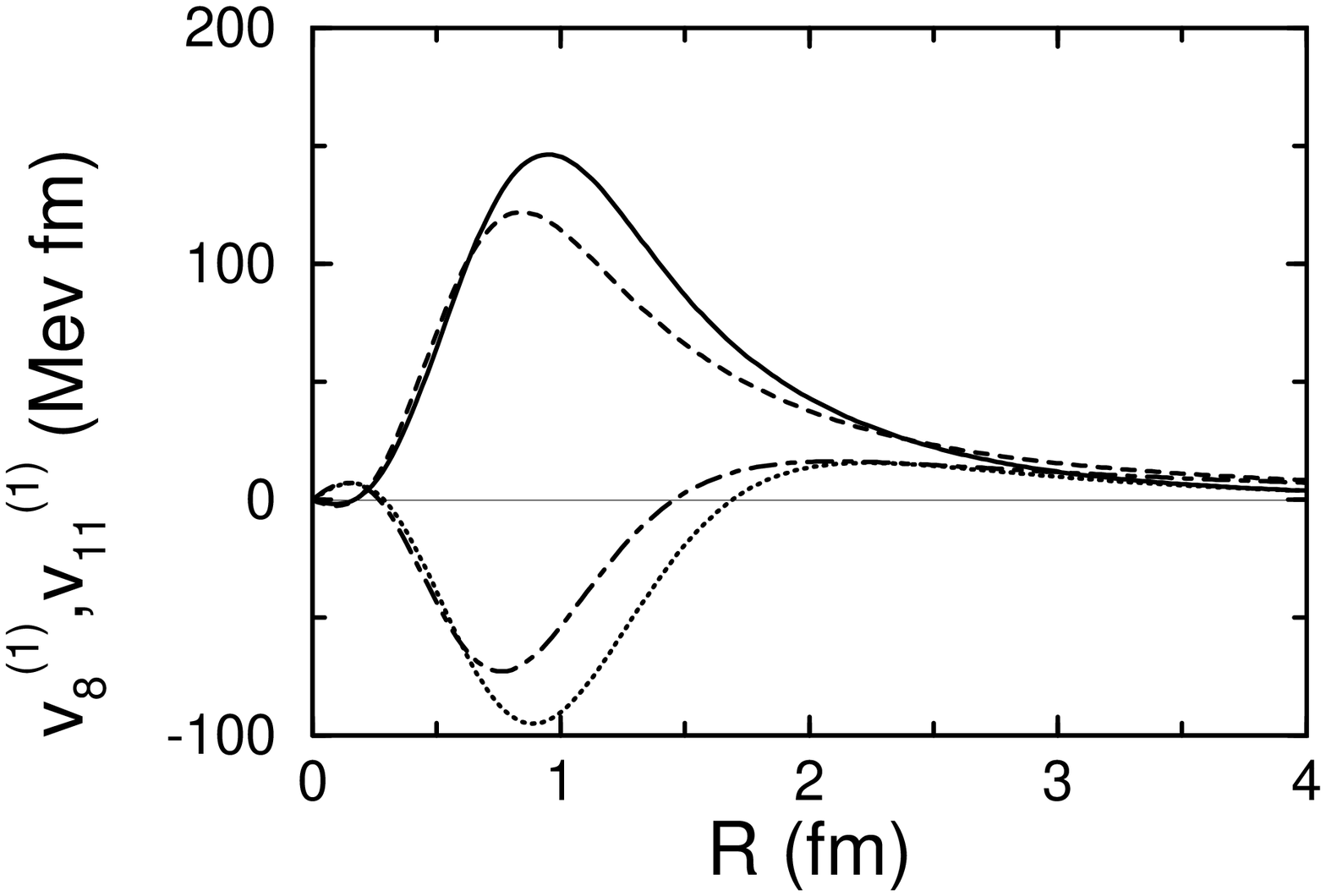}}
\caption{Solid line: $v_{8}^{(1)}$ with pion mass term; dashed line:
$v_{8}^{(1)}$, zero pion mass; dotted: $v_{11}^{(1)}$, with pion mass
term; dot-dashed: $v_{11}^{(1)}$, with zero pion mass.\label{fig:14}}
\end{figure}
\begin{figure}
\epsfysize=5cm
\centerline{\epsffile{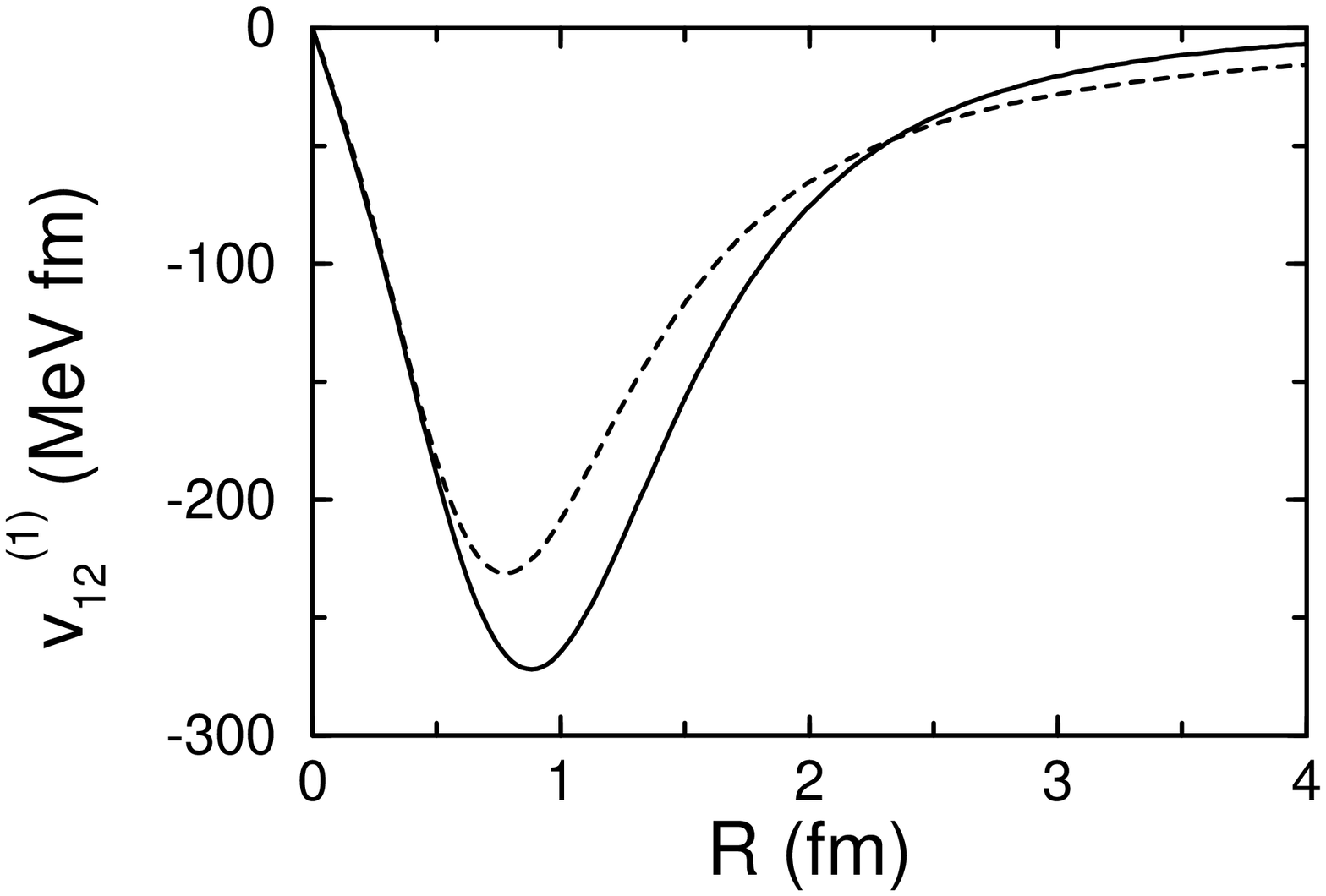}}
\caption{Solid line: $v_{12}^{(1)}$, with pion mass term; dashed line:
$v_{12}^{(1)}$, zero pion mass.\label{fig:15}}
\end{figure}
$v^{(1)}_9$, $v^{(1)}_{10}$ and $v^{(1)}_{13}$ are very large
parity violating terms that decay slowly with $R$ (see Figs.~16--17).
\begin{figure}
\epsfysize=5cm
\centerline{\epsffile{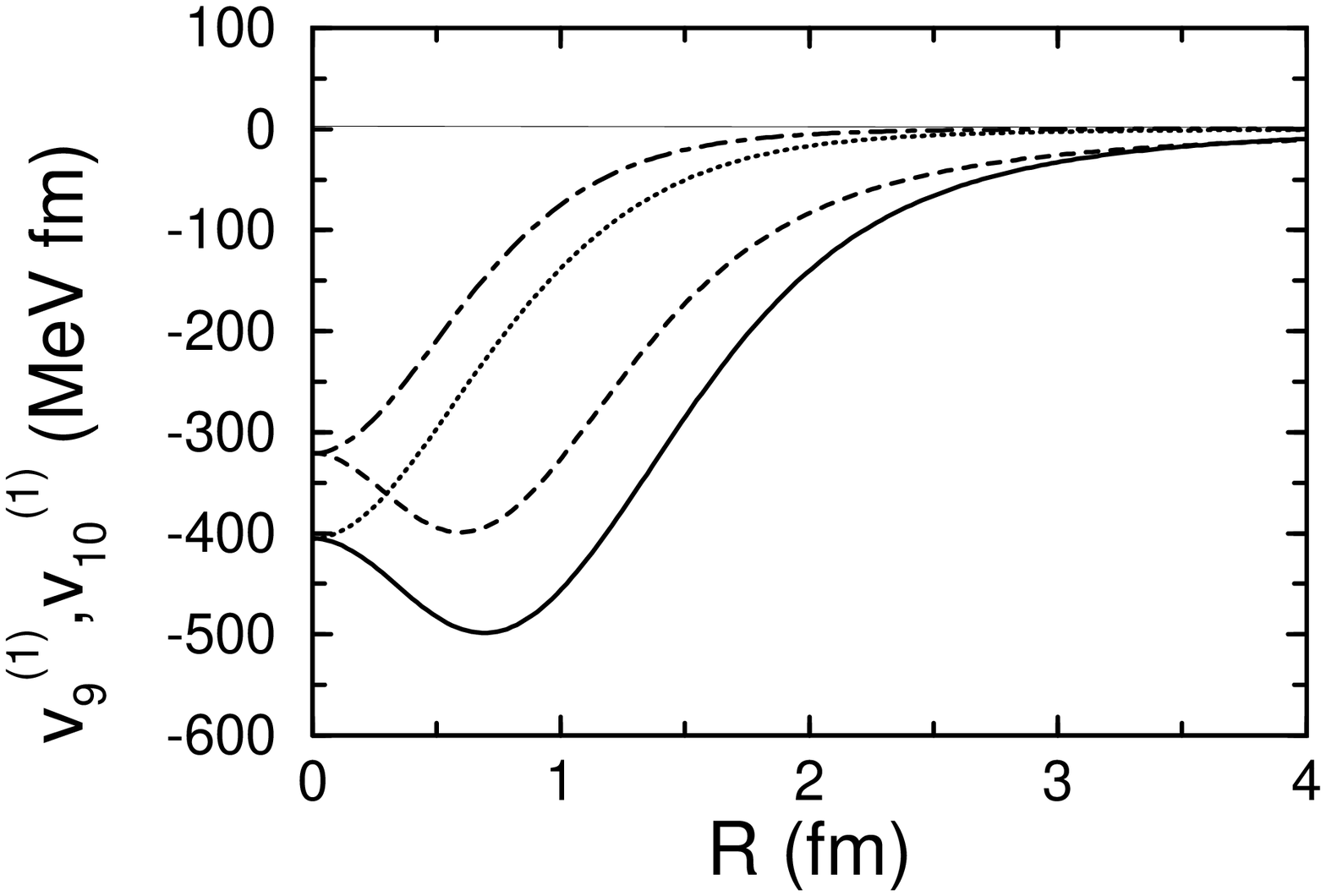}}
\caption{Solid line: $v_{9}^{(1)}$, with pion mass term; dashed line:
$v_{9}^{(1)}$, zero pion mass; dotted: $v_{10}^{(1)}$, with pion mass
term; dot-dashed: $v_{10}^{(1)}$, with zero pion mass.
 These terms are odd under parity.\label{fig:16}}
\end{figure}
\begin{figure}
\epsfysize=5cm
\centerline{\epsffile{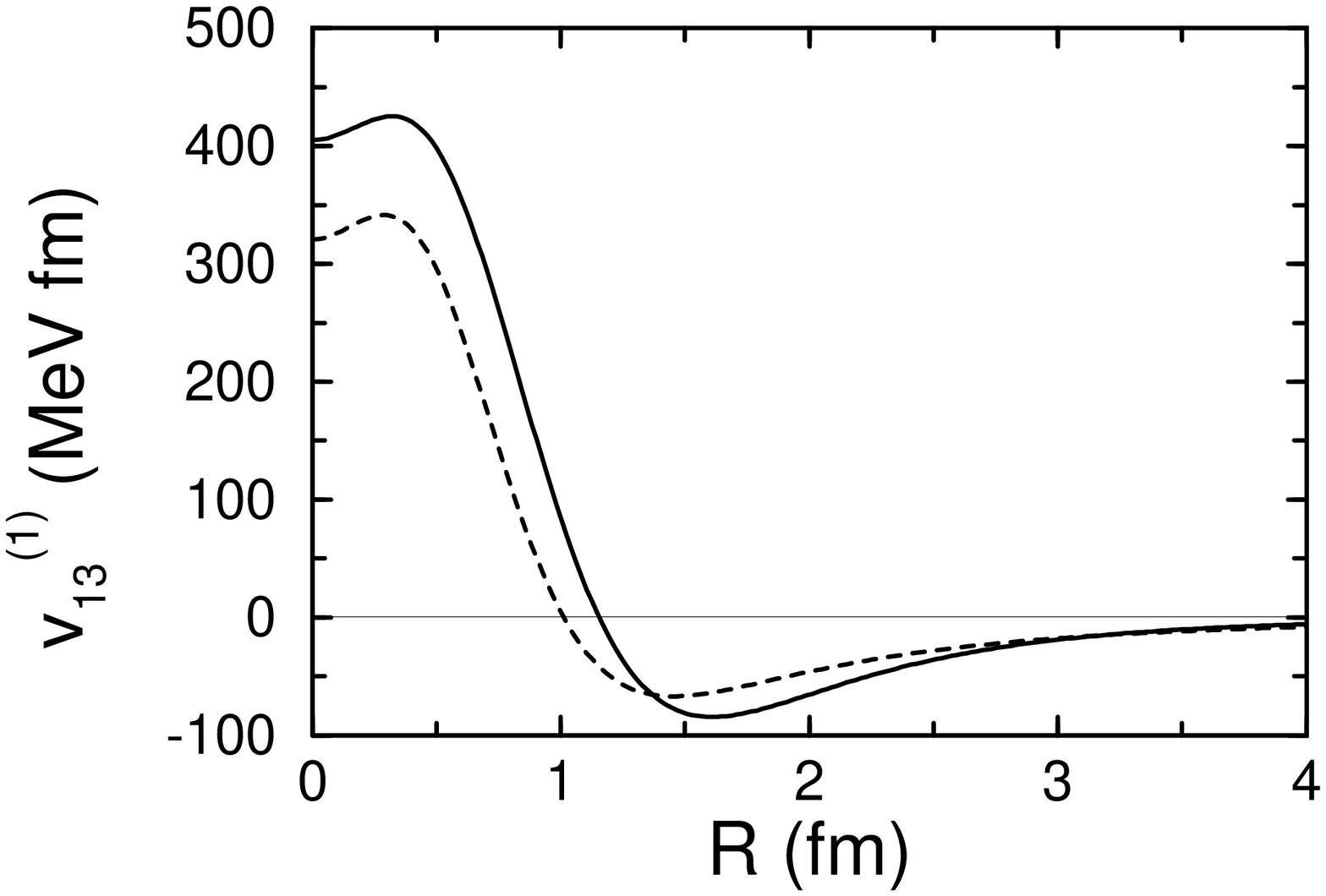}}
\caption{Solid line: $v_{13}^{(1)}$, with pion mass term; dashed line:
$v_{13}^{(1)}$, zero pion mass. This term is odd under parity.\label{fig:17}}
\end{figure}
They are troublesome. All of the $w^{(1)}_i$ terms (quartic in $C$)
are essentially
zero outside 1 fm.  Of the Greek letter terms, $\phi^{(1)}_2$, $\phi^{(1)}_4$
and $\phi^{(1)}_5$ are parity conserving and large, while $\phi^{(1)}_6$ and
$\phi^{(1)}_7$ are large, slowly decreasing with $R$ and parity violating
(see Figs.~18--20).
\begin{figure}
\epsfysize=5cm
\centerline{\epsffile{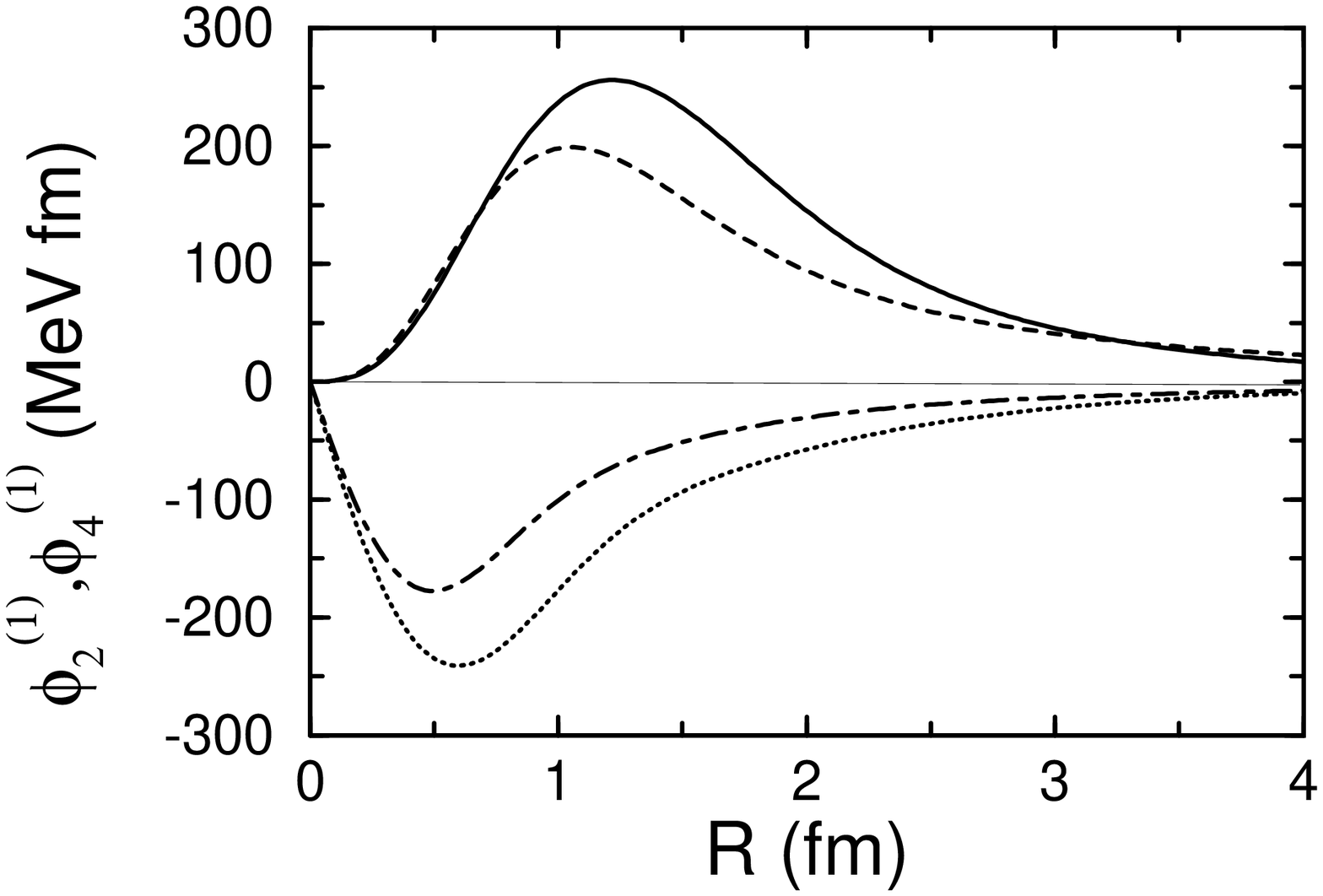}}
\caption{Solid line: $\phi_{2}^{(1)}$, with pion mass term; dashed line:
$\phi_{2}^{(1)}$, zero pion mass; dotted: $\phi_{4}^{(1)}$, with pion mass
term; dot-dashed: $\phi_{4}^{(1)}$, with zero pion mass.\label{fig:18}}
\end{figure}
\begin{figure}
\epsfysize=5cm
\centerline{\epsffile{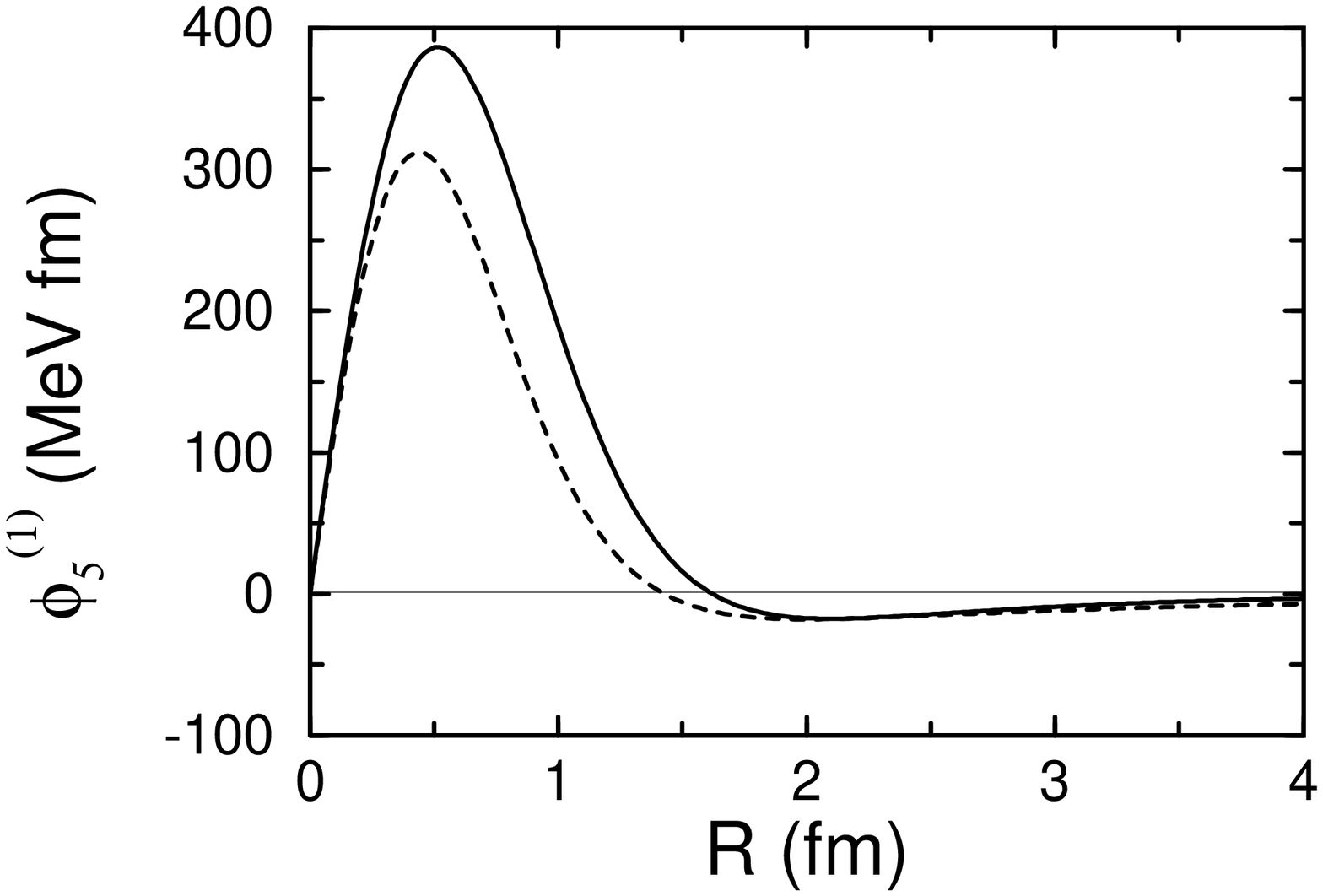}}
\caption{Solid line: $\phi_{5}^{(1)}$, with pion mass term; dashed line:
$\phi_{5}^{(1)}$, zero pion mass.\label{fig:19}}
\end{figure}
\begin{figure}
\epsfysize=5cm
\centerline{\epsffile{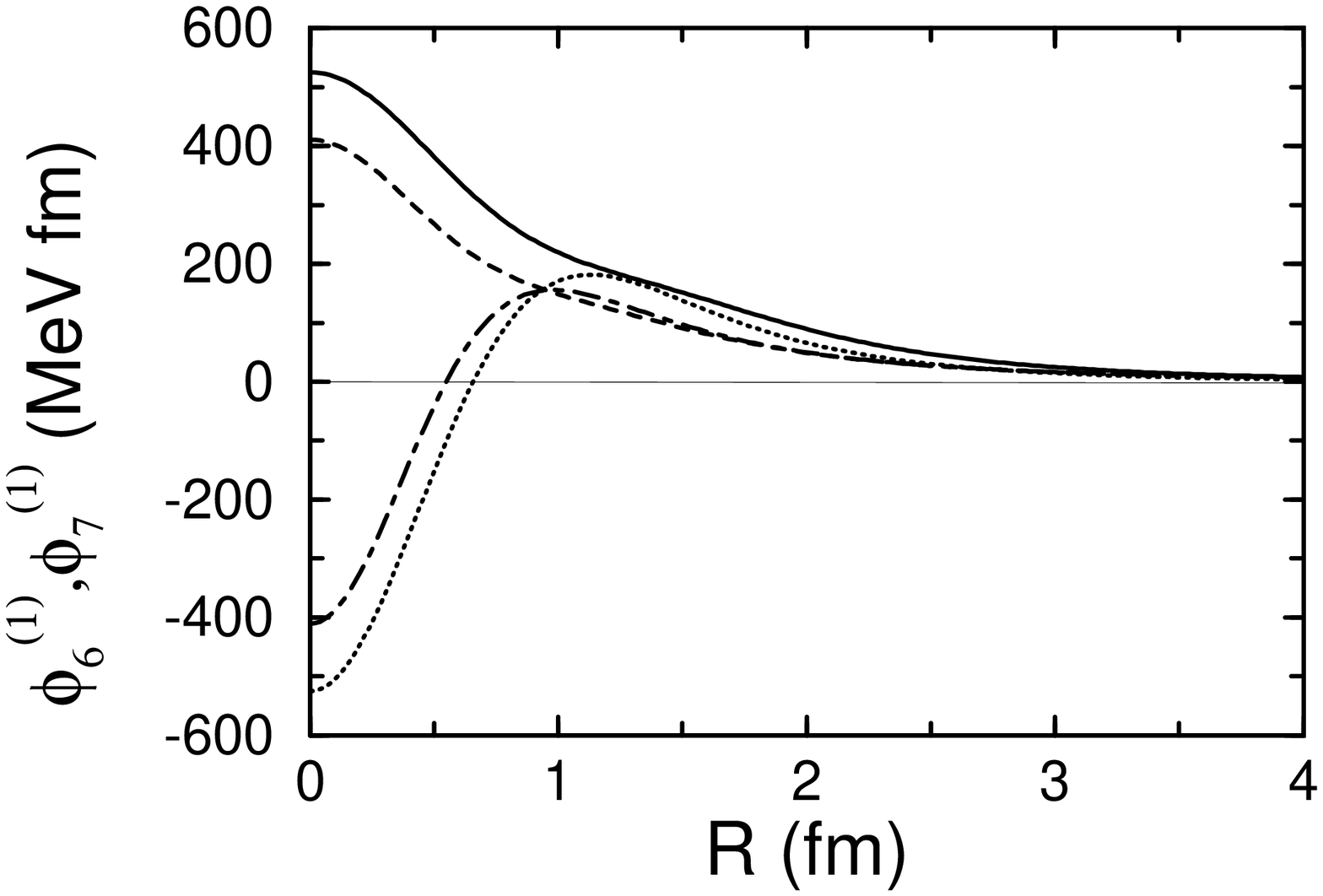}}
\caption{Solid line: $\phi_{6}^{(1)}$ with pion mass term; dashed line:
$\phi_{6}^{(1)}$, zero pion mass; dotted: $\phi_{7}^{(1)}$, with pion mass
term; dot-dashed: $\phi_{7}^{(1)}$, with zero pion mass.\label{fig:20}}
\end{figure}
All of the $\eta^{(1)}_i$ terms (quartic in $C$) are very
small outside of 1 fm.

\subsection{Iso-rotational Coupling}

Finally we consider $K_2$ with $\bbox{t}_{1} = \bbox{R}_{t}(A_{1})$ and
$\bbox{t}_{2} =\bbox{R}_{t}(A_{2})$.  We find $u^{(0)}_1$, $u^{(0)}_2$,
$u^{(0)}_3$, $v^{(0)}_1$, $v^{(0)}_2$, $v^{(0)}_3$,
$v^{(0)}_5$, $v^{(0)}_6$, $v^{(0)}_7$, and $v^{(0)}_8$
all to be very small.
\begin{figure}
\epsfysize=5cm
\centerline{\epsffile{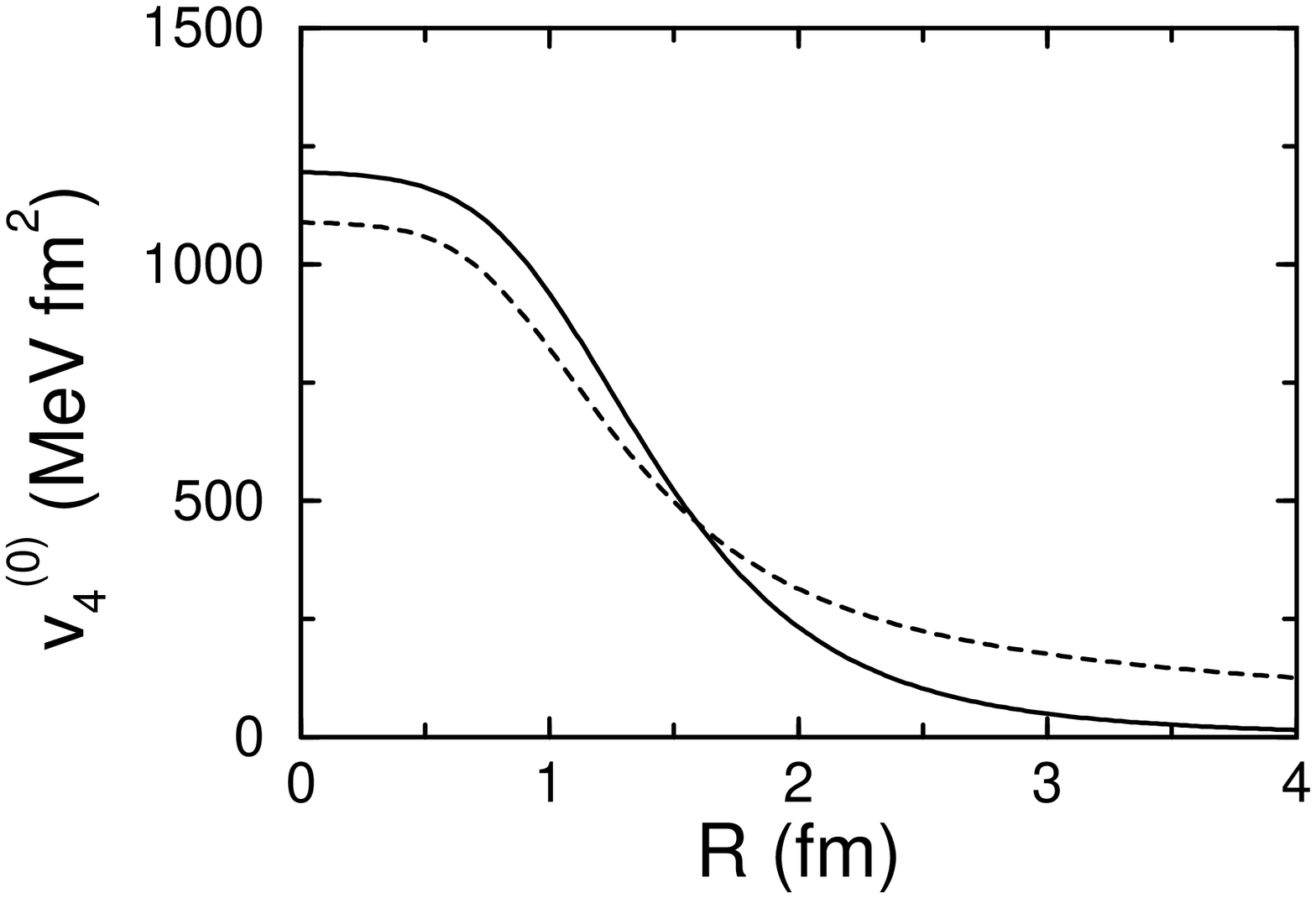}}
\caption{Solid line : $v_{4}^{(0)}$ with pion mass term; dashed line:
$v_{4}^{(0)}$, zero pion mass.\label{fig:21}}
\end{figure}
\begin{figure}
\epsfysize=5cm
\centerline{\epsffile{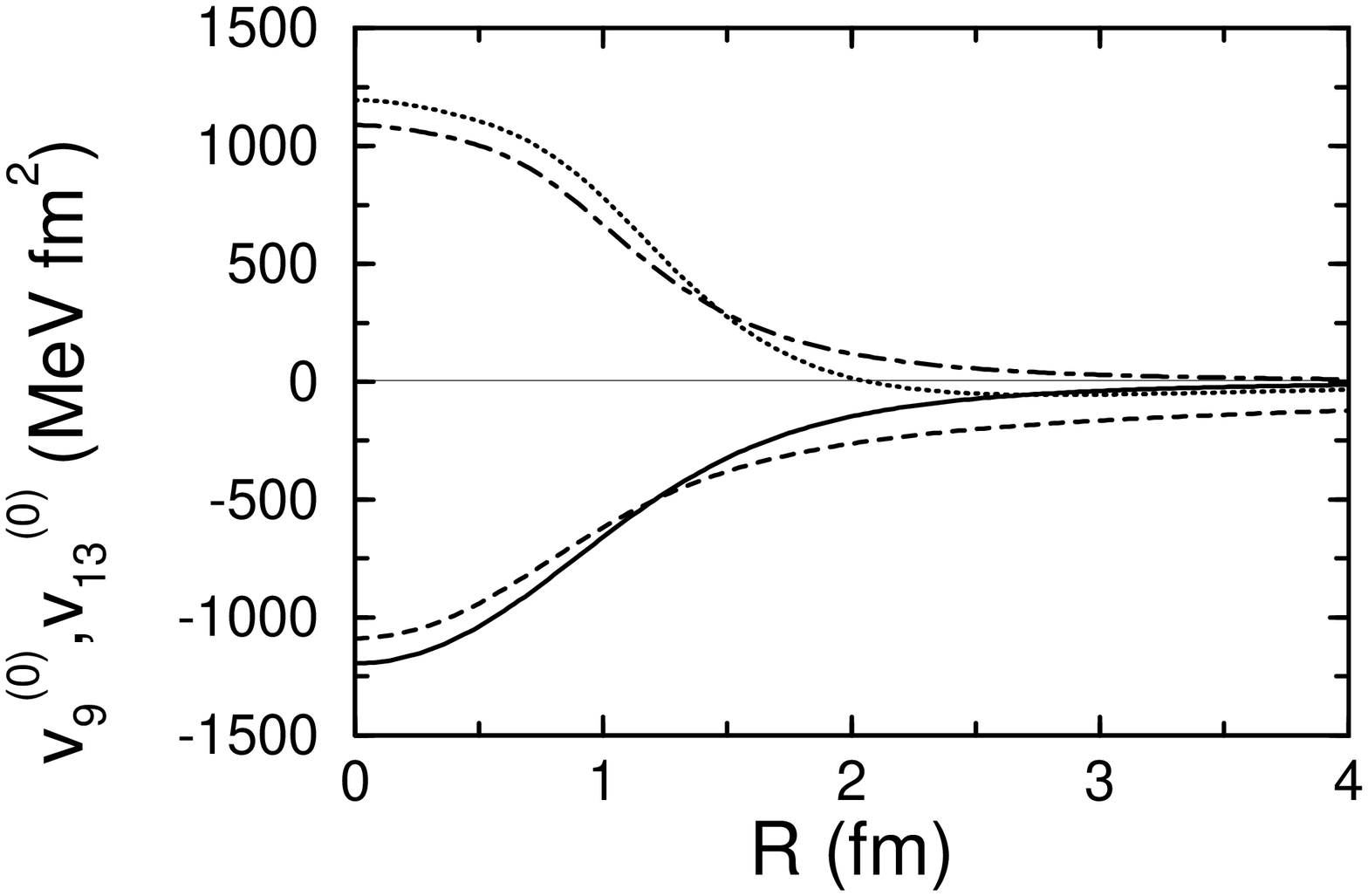}}
\caption{Solid line: $v_{9}^{(0)}$, with pion mass term; dashed line:
$v_{9}^{(0)}$, zero pion mass; dotted: $v_{13}^{(0)}$, with pion mass
term; dot-dashed: $v_{13}^{(0)}$, with zero pion mass.\label{fig:22}}
\end{figure}
We find $v^{(0)}_4$, $v^{(0)}_9$, $v^{(0)}_{10}$ and $v^{(0)}_{13}$
to be large (see Figs.~21--22, $v^{(0)}_{10}$ is not plotted
since it is equal to $v^{(0)}_{9}$ as expected from permutation symmetry).
They also decay
very slowly for large $R$.  Analysis shows that they go like $1/R$ for
large $R$ in the case of zero pion mass.
The remaining terms in $v$ and all the terms in $w$ (quartic
in $C$) are small outside 1 fm.
$\phi^{(0)}_3$, $\phi^{(0)}_6$ and $\phi^{(0)}_7$ are very
large and exhibit considerable sensitivity to pion mass (see Fig.~23,
$\phi^{(0)}_7$ is not plotted since it is equal to $\phi^{(0)}_{6}$
as expected from permutation symmetry).
\begin{figure}
\epsfysize=5cm
\centerline{\epsffile{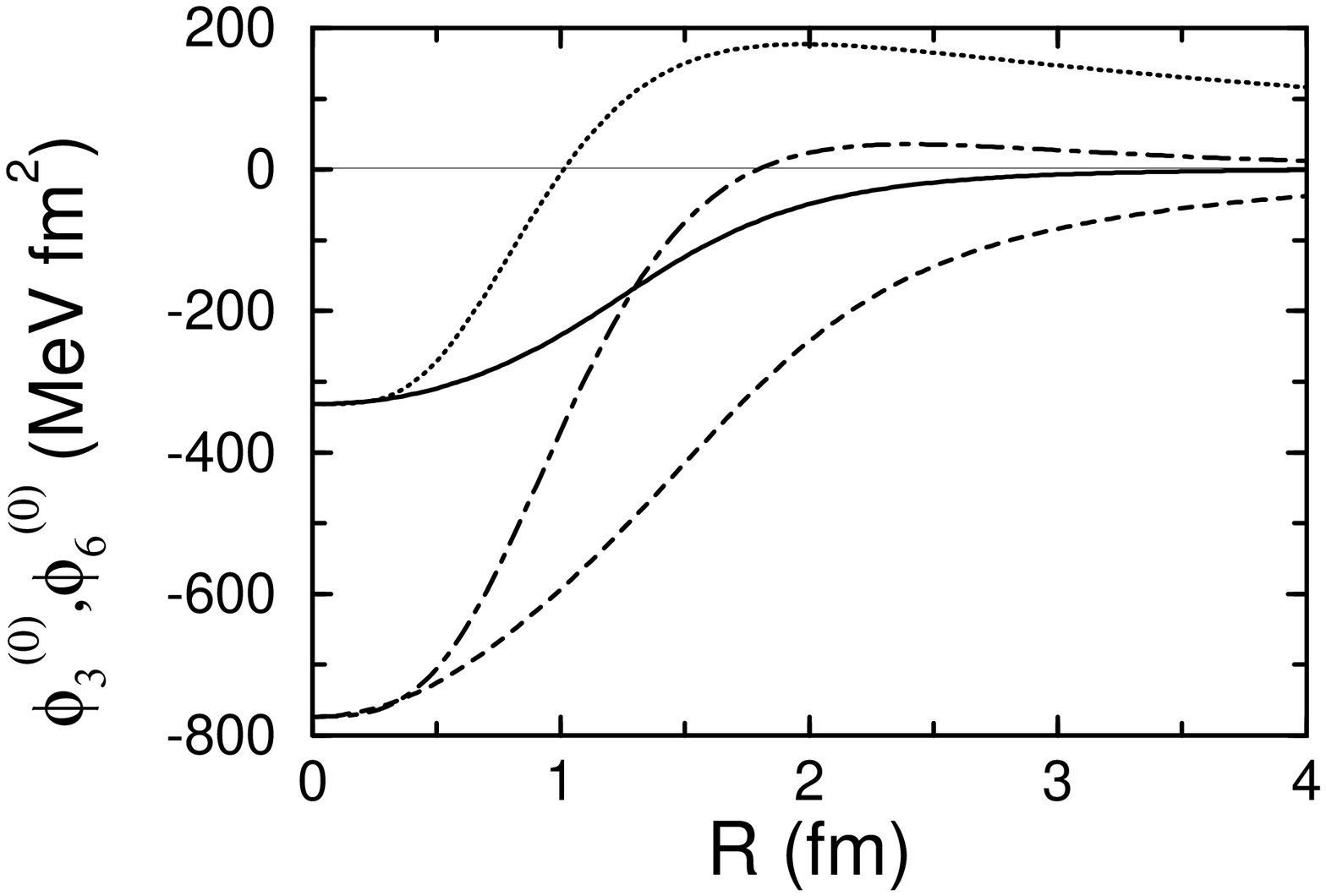}}
\caption{Solid line: $\phi_{3}^{(0)}$, with pion mass term; dashed line:
$\phi_{3}^{(0)}$, zero pion mass; dotted: $\phi_{6}^{(0)}$, with pion mass
term; dot-dashed: $\phi_{6}^{(0)}$, with zero pion mass.\label{fig:23}}
\end{figure}
and $\phi^{(0)}_5$ are large and parity violating (see Fig.~24,
only $\phi^{(0)}_4$ is plotted since  $\phi^{(0)}_5=-\phi^{(0)}_4$,
thus violating permutation symmetry as well).
\begin{figure}
\epsfysize=5cm
\centerline{\epsffile{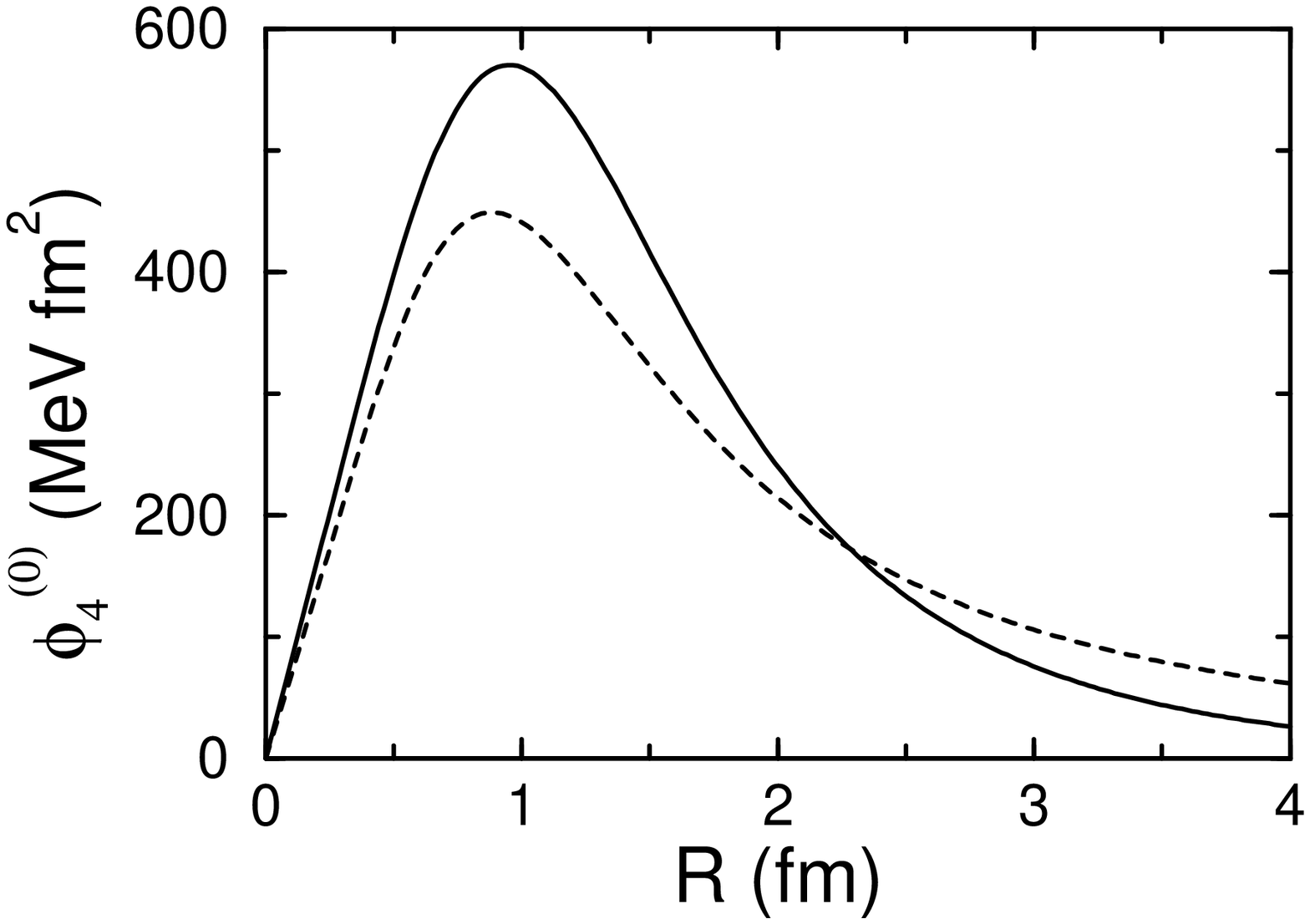}}
\caption{Solid line: $\phi_{4}^{(0)}$ with pion mass term; dashed line:
$\phi_{4}^{(0)}$, zero pion mass. This term is odd under parity.
\label{fig:24}}
\end{figure}
Again all the quartic terms, $\eta$, are small.

\section{Conclusions}
\label{sec:5}

In this paper we present a systematic development of the
kinetic energy of the $B=2$ Skyrmion system starting from the product
ansatz. Although this ansatz has serious shortcomings, it provides
a useful starting point for a first orientation to the velocity dependent
terms in the Skyrme energy.  It is particularly easy to identify the
collective degrees of freedom in the product ansatz and therefore to
make them time dependent.  Furthermore the problems of the product
ansatz in calculating the static energy of the $B=2$ system, namely the
subtraction of large numbers to get a small one, are not present in
the kinetic energy.

The full kinetic energy of the $B=2$ Skyrmion system is a rich and complex
expression.  There are three dynamical velocities that enter, the
relative velocity of the two Skyrmions and the rotational velocity of each.
We have not considered the trivial overall translational velocity of the
entire system.
The kinetic energy is
bilinear in the velocities and all possible couplings
among the three velocity vectors
enter, so that the mass matrix is nine by nine.
The kinetic energy is a scalar function of the velocities, the relative
separation vector of the two Skyrmions and of the vector that
specifies their relative iso-orientation.
It is convenient to classify the dependence on iso-orientation in terms
of irreducible tensors of definite degree in the relative
orientation, $C$.  Call that degree $\sigma$. (It is in fact
the $O(4)$ quantum number, see Appendix).
We find terms with $\sigma=0,2,4$.
The terms with $\sigma=0$ correspond to no iso-orientational dependence
and lead to iso-spin independent contributions to the kinetic energy
of two nucleons.  The terms with $\sigma=2$ lead to kinetic energy terms for
two nucleons that have an iso-spin dependence of the form $\bbox{\tau}_1
\cdot \bbox{\tau}_2$. We find that, in general,
the $\sigma=2$ terms are smaller
for the two Skyrmion system than the $\sigma=0$ terms.
The $\sigma=4$ terms are
smaller still and have vanishing matrix element in the two nucleon system,
but will couple with deltas.

We find that there are some non-zero terms in the expansion of the kinetic
energy that are not even under parity and are not symmetric under
interchange of the two Skyrmions.  These terms are pathological
artifacts of the product ansatz that would not be present in an exact
calculation. Some of these terms are quite large and fall off very
slowly with the separation of the Skyrmions, particularly in the case of
zero pion mass.  These large terms are particularly worrisome and give
an ominous sense of the limits of the product ansatz calculation.

Our major purpose is to give a full and systematic discussion of the
two Skyrmion kinetic energy to serve as a foundation for any discussion
of non-static terms in the two Skyrmion and two nucleon interaction.
Here we only present the full algebraic development of the kinetic
energy and some result for the largest terms with their dependence
on the relative separation of the Skyrmions. There are many applications
of these results that come to mind.  For example the leading non-static
contribution to the energy of two nucleons is the spin-orbit interaction.
It arises from the terms in the kinetic energy that couple the relative
velocity and the total rotational velocity. To go from these terms
to a nucleon-nucleon spin-orbit interaction in a Hamiltonian
require that the mass matrix be inverted before projecting on
nucleon states. We have recently shown
that doing this yields an iso-spin independent spin-orbit interaction of
the correct magnitude and sign as well as explaining the relatively
small size of the iso-spin dependent spin-orbit interaction \cite{SpinOrbit}.
There are many more non-static terms in the nuclear force that invite study.
There is also the opportunity to go beyond the product ansatz.
Here the problem of defining collective variables comes up.  We are
studying this difficult issue.  A starting point, pursued with considerable
success in the static energy studies, is to start with the kinetic energy
in the two Skyrmion channels with reflection symmetry.  This is easily done
in the product ansatz and we are studying the advantages of this approach
for an exact treatment.

In summary, we have
studied  the kinetic energy of the $B=2$ Skyrmion
system in the product ansatz in a systematic expansion in
the collective velocitites.
This is a first step in adding non-static
corrections to the recent success for the static nucleon-nucleon interaction
\cite{WA}
calculated in an approach based in non-perturbative QCD.
\vspace{3mm}

This research was supported in part by a grant from the
U. S. National Science Foundation.

\appendix
\section{Irreducible Representations of the $O(4)$ in Relative Iso-rotation}
The expansion of the kinetic energy is a sum of terms quadratic in the
translational and isorotational velocities.  The coefficients
depend only on  the relative isorotation $C=A_{1}^{\dagger}A_{2}$. Since
$C=c_{4}+i\bbox{c}\cdot\bbox{\tau}$ is unitary, the $c$'s satisfy
\begin{equation}
c_{4}^2+\bbox{c}^2=1.
\end{equation}
The operators
\begin{eqnarray}
&& K_{i}=-i\epsilon_{ijk}c_{j}\frac{\partial}{\partial c_{k}} \nonumber \\ &&
D_{i}=-i(c_{4}\frac{\partial}{\partial c_{i}}-c_{i}
\frac{\partial}{\partial c_{4}})
\end{eqnarray}
form an $O(4)$ algebra. The $K_{i}$'s generate an  $SO(3)$ subalgebra.
The Casimir operator of $O(4)$ is $C_{2}(O(4))=\bbox{K}^2+\bbox{D}^2$
and its eigenvalue is $\sigma(\sigma+2)$.
The Casimir operator of the
$SO(3)$ subalgebra is  $C_{2}(O(3))=\bbox{K}^2$ with eigenvalue $k(k+1)$.
It is easy to show that the following polynomials in the four components
of $C$, spherical harmonics of $O(4)$,
carry irreducible representations of the group chain $O(4)\supset O(3)$:

\begin{flushleft}
$\sigma=2$
\end{flushleft}

\begin{equation}
T^{(0)}=\bbox{c}^2-3c_{4}^2
\end{equation}

\begin{equation}
T^{(1)}_{i}=c_{i}c_{4}
\end{equation}

\begin{equation}
T^{(2)}_{ij}=c_{i}c_{j}-\frac{1}{3}\bbox{c}^2\delta_{ij}
\end{equation}

\begin{flushleft}
$\sigma=4$
\end{flushleft}

\begin{equation}
F^{(0)}=(\bbox{c}^2)^2-10c_{4}^2\bbox{c}^2+5c_{4}^4
\end{equation}

\begin{equation}
F^{(1)}_{i}=c_{i}\bbox{c}^2 c_{4}-\frac{5}{3}c_{i}c_{4}^3
\end{equation}

\begin{equation}
F^{(2)}_{ij}=c_{i}c_{j}c_{4}^2-\frac{1}{7}c_{i}c_{j}\bbox{c}^2
-\frac{1}{3}\bbox{c}^2 c_{4}^2\delta_{ij}+\frac{1}{21}(\bbox{c}^2)^2\delta_{ij}
\end{equation}

\begin{equation}
F^{(3)}_{ijk}=c_{i}c_{j}c_{k}c_{4}-\frac{1}{5}
(c_{i}\delta_{jk}+c_{j}\delta_{ik}+c_{k}\delta_{ij})\bbox{c}^2 c_{4}
\end{equation}

\begin{eqnarray}
F^{(4)}_{ijkl}&=&c_{i}c_{j}c_{k}c_{l}
-\frac{1}{7}\bbox{c}^2(c_{i}c_{j}\delta_{kl}+c_{i}c_{k}\delta_{jl}+
c_{i}c_{l}\delta_{jk}+\nonumber \\ &&
c_{j}c_{k}\delta_{il}+
c_{j}c_{l}\delta_{ik}+c_{k}c_{l}\delta_{ij}) \nonumber \\ &&
+\frac{1}{35}(\delta_{ij}\delta_{kl}+\delta_{ik}\delta_{jl}
+\delta_{il}\delta_{jk})(\bbox{c}^2)^2
\end{eqnarray}

The $T^{(k)}$ polynomials are quadratic in $C$ and carry the
irreducible representations $(\sigma=2,k)$ of $O(4)\supset O(3)$.
They are linear  combinations of the matrix elements
of the Wigner function $D^{(1)}(C)$.
The $F^{(k)}$ polynomials are quartic in $C$ and carry the
irreducible representations $(\sigma=4,k)$ of $O(4)\supset O(3)$.
$F$ functions are linear combinations of the matrix elements of
the Wigner function $D^{(2)}(C)$ and their matrix elements between
nucleon-nucleon states are therefore zero.

In terms of $T$ and $F$ polynomials in $C$, we use the following notation
for any vectors $\bbox{A},\bbox{B},\bbox{C},\bbox{D}$:
\begin{equation}
\begin{array}{l}
T^{(1)}(\bbox{A})=T^{(1)}_{i}A_{i}, \\
T^{(2)}(\bbox{A},\bbox{B})=T^{(2)}_{ij}A_{i}B_{j},
\end{array}
\end{equation}
and
\begin{equation}
\begin{array}{l}
F^{(1)}(\bbox{A})=F^{(1)}_{i}A_{i}, \\
F^{(2)}(\bbox{A},\bbox{B})=F^{(2)}_{ij}A_{i}B_{j}, \\
F^{(3)}(\bbox{A},\bbox{B},\bbox{C})=F^{(3)}_{ijk}A_{i}B_{j}C_{k}, \\
F^{(4)}(\bbox{A},\bbox{B},\bbox{C},\bbox{D})=F^{(4)}_{ijkl}A_{i}B_{j}C_{k}D_{l}.
\end{array}
\end{equation}

\end{document}